\documentclass[%
 preprint,
 amsmath,amssymb,
 aps,
showkeys ]{revtex4}

\usepackage{graphicx}
\usepackage{dcolumn}
\usepackage{bm}
\usepackage{float}
\usepackage{color}

\begin{document}


\title{Disordered Hyperuniform Solid State Materials}

\author{Duyu Chen}
\email[correspondence sent to: ]{duyu@ucsb.edu}
\affiliation{Materials Research Laboratory, University of California, Santa Barbara, California 93106, United States}
\author{Houlong Zhuang}
\email[correspondence sent to: ]{zhuanghl@asu.edu}
\affiliation{Mechanical and Aerospace Engineering, Arizona State
University, Tempe, AZ 85287}
\author{Mohan Chen}
\affiliation{HEDPS, Center for Applied Physics and Technology, College of Engineering and School of Physics, Peking University, Beijing 100871, People’s Republic of China}
\author{Pinshane Y. Huang}
\affiliation{Department of Materials Science and Engineering, University of Illinois Urbana-Champaign, Urbana, IL 61801, United States}
\author{Vojtech Vlcek}
\affiliation{Department of Chemistry and Biochemistry,
University of California, Santa Barbara, CA 93106, United States}
\author{Yang Jiao}
\email[correspondence sent to: ]{yang.jiao.2@asu.edu}
\affiliation{Materials Science and Engineering, Arizona State
University, Tempe, AZ 85287} \affiliation{Department of Physics,
Arizona State University, Tempe, AZ 85287}



\date{\today}

\begin{abstract}

Disordered hyperuniform (DHU) states are recently discovered exotic states of condensed matter. DHU systems are similar to liquids or glasses in that they are statistically isotropic and lack conventional long-range translational and orientational order. On the other hand, they completely suppress normalized infinite-wavelength density fluctuations like crystals, and in this sense possess a hidden long-range order. Very recently, there are several exciting discoveries of disordered hyperuniformity in solid-state materials, including amorphous carbon nanotubes, amorphous 2D silica, amorphous graphene, defected transition metal dichalcogenides, defected pentagonal 2D materials, and medium/high-entropy alloys. It has been found the DHU states of these materials often possess a significantly lower energy than other disorder models, and can lead to unique electronic and thermal transport properties, which resulted from mechanisms distinct from those identified for their crystalline counterparts. For example, DHU states can enhance electronic transport in 2D amorphous silica; DHU medium/high-entropy alloys realize the Vegard's law, and possess enhanced electronic band gaps and thermal transport at low temperatures. These unique properties open up many promising potential device applications in optoelectronics and thermoelectrics. Here, we provide a focused review on these important new developments of hyperuniformity in solid-state materials, taking an applied and ``materials'' perspective, which complements the existing reviews on hyperuniformity in physical systems and photonic materials. Future directions and outlook are also provided, with a focus on the design and discovery of DHU quantum materials for quantum information science and engineering.


\end{abstract}


\keywords{Disorder Hyperuniformity, Nanotubes, 2D Materials, High/Medium-Entropy Alloys, Electronic Structures, Thermal Properties}


\maketitle


\newpage
\section{Introduction}


Solid-state materials, in either bulk or thin-film forms, inevitably contain disorder of various kinds \cite{kittel1996introduction}. At finite temperatures, the otherwise perfect crystal lattice is constantly distorted by the thermal motions of the atoms. Besides this normal ``thermal disorder'', which can be well understood in the framework of statistical thermodynamics, there are a wide spectrum of ``frozen-in disorder'', including point defects, dislocations, grain boundaries and surfaces, to name but a few. The physical properties of the materials, such as electronic band structures, density of states, electrical and thermal transport properties, heat capacity, magnetic and optical properties, etc., strongly depend on the concentration, spatial distribution and interactions among the defects, which have been the subjects of intensive studies in condensed matter and solid-state physics \cite{baranovskii2017charge, lee1985disordered, dugdale, kramer1993localization, sheng2007introduction}. Traditionally, it was believed that defects may have negative impacts on material properties, while recently it has been shown that novel desirable material performance can be achieved by proper defect engineering \cite{lin2016defect, zhang2019defect, shi2021defect}. For example, it has been demonstrated that quantum emitters can be realized by introducing defects into two-dimensional (2D) hexagonal boron nitride ($h$-BN) \cite{tran2016quantum}.

Another type of ``frozen-in disorder'' is typically found in amorphous or ``glassy'' states of the materials, where the disorder is so significant that the material cannot be considered as perturbed or defected states of otherwise perfect crystals \cite{binder2011glassy, angell2000relaxation}. Examples include correlated disorder, which resulted mostly from fast cooling, in metallic glasses \cite{wang2004bulk} and amorphous silicon \cite{ref28, ref29};  compositional disorder in alloys due to entropy-driven mixing \cite{porter2009phase}; and spin disorder in randomly diluted magnetic materials (i.e., spin glasses) \cite{mezard1987spin, mezard1984nature}. Compared to their crystalline counterparts, the amorphous materials lack conventional long-range translational and orientational order, e.g., there are no Bragg peaks in their scattering intensity. On the other hand, they can possess significant short-range order manifested by clearly structured pair-distributions \cite{ref29, Ch21}, as a result of local chemical bonding and electronic structures. Importantly, the amorphous materials generally preserve the ``isotropy'' property of the high-symmetry liquid phase, which is highly desirable for many applications (e.g., the window glass) \cite{binder2011glassy}.


The nature of the aforementioned ``frozen-in disorder'' generally depends on the specific material systems of interest. An important yet very challenging step to develop a comprehensive theory for amorphous materials is the systematic and rigorous mathematical description of such disorder and the understanding of its effects on different material properties. To this end, a variety of stochastic morphological models have been developed including random point processes, random sets and random field models \cite{jeulin2021morphological}. Statistical descriptors such as the hierarchy of $n$-point correlation functions \cite{To02a}, which in their most general settings quantify the probability of occurrence of a unique combination of local states at specific locations in the material of interest, have been applied in a diverse class of disordered condensed matter systems. Moreover, percolation models \cite{sahini1994applications, torquato2013effect} have been widely employed to understand the structural and transport properties of amorphous materials.


\begin{figure*}[ht]
\begin{center}
$\begin{array}{c}\\
\includegraphics[width=0.95\textwidth]{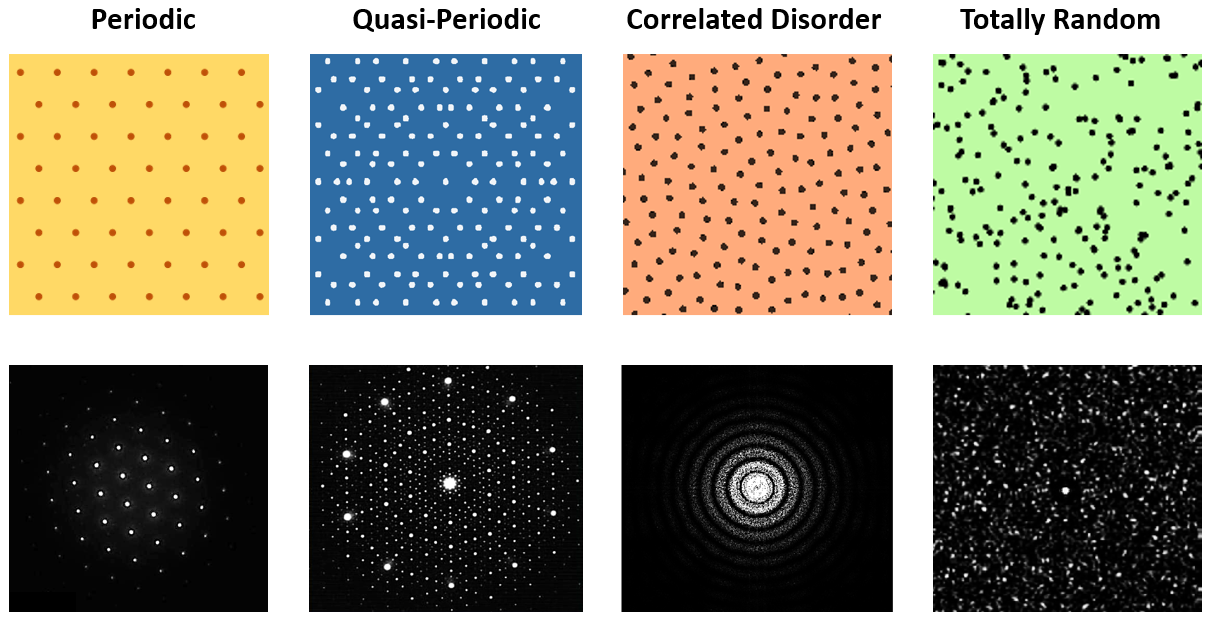}
\end{array}$
\end{center}
\caption{Illustration of point configurations with different degrees of disorder. From left to right: periodic configuration (crystal), quasi-periodic configuration (quasi-crystal), correlated disordered configuration (disordered material), totally random configuration (ideal gas). The lower panels show the static structure factor $S({\bf k})$ for the corresponding point configurations (see definition in Sec. II).} \label{fig1}
\end{figure*}

Recently, a new framework for quantifying and classifying disorder has been proposed based on the concept of ``hyperuniformity'' \cite{To18a}. The notion of hyperuniformity was first coined by Torquato and Stillinger \cite{To03} in the development of universal order metrics to distinguish point configurations with different degrees of order, including periodic patterns, quasi-periodic patterns, correlated disordered patterns, and totally random patterns (see illustrations in Fig. \ref{fig1}). These configurations respectively correspond to distributions of atoms in crystals, quasicrystals, amorphous materials and ideal gases. Yet, the notion of hyperuniformity can be readily generalized to, e.g., distribution of chemical species in alloys (i.e., chemical but not spatial disorder). There are also no fundamental limitations of hyperuniformity to classical particles (and classical properties) and it may offer a new apparatus to capturing correlations of quantum states (i.e., probability amplitudes, densities, density matrices, etc). The generalization of hyperuniformity to the aforementioned systems will be discussed in detail in Sec. II.B.

The definition of hyperuniformity involves the measurement of density fluctuations in the system across different length scales (see Fig. \ref{fig2}). Specifically, one can consider placing spherical observation windows $\Omega(R)$ of radius $R$ at random locations in the system. As the window moves from one (randomly selected) location to another, the number of points (atoms) falling into the window $N(R)$ also fluctuates. The variance $\sigma^2_N(R)$ associated with $N(R)$ is also a function of window radius $R$. For crystals and quasicrystals, as the windows moves in the system, one can imagine that the ``bulk'' portion enclosed the window would not change, especially for large windows, due to the long-range translational and/or orientational order in the systems. Thus, the contributions to the number fluctuations only come from the ``surface layer'' of the window, leading to $\sigma^2_N(R) \sim R^{d-1}$ (where $d$ is the dimensionality of the system and $R^{d-1}$ describes the scaling of the surface area of a spherical window). On the other hand, it is well known that for ideal gases, which is a totally un-correlated system, $\sigma^2_N(R) \sim R^{d}$.


A hyperuniform system is one whose density fluctuations are strongly suppressed on large length scales and possesses vanishing normalized number variance for infinite windows, i.e., $\lim_{R\rightarrow\infty} \sigma^2_N(r)/R^d = 0$. Thus, all crystals and quasicrystals are hyperuniform by definition. Interestingly, certain exotic correlated disordered systems, which can be in both equilibrium and non-equilibrium settings, and come in both quantum-mechanical and classical varieties, are also found to be hyperuniform. These {\it disordered hyperuniform} (DHU) systems are similar to liquids or glasses in that they are statistically isotropic and lack conventional long-range translational order. Yet they possess a {\it hidden long-range order}, manifested as complete suppression of large-scale normalized density fluctuations like in crystals. Examples of DHU systems include the density fluctuations in early
universe \cite{ref3}, disordered jammed packing of hard particles
\cite{ref4, ref5, ref6, ref7, atkinson2012maximally, Za11c, Za11d}, certain exotic classical ground
states of many-particle systems \cite{ref8, ref9, ref10, ref11,
ref12, ref13, ref14, ref15}, jammed colloidal systems \cite{ref16,
ref17, ref18, ref19}, driven non-equilibrium systems \cite{ref20,
ref21, ref22, ref23, salvalaglio2020hyperuniform}, certain quantum ground states \cite{ref24, ref25}, avian photoreceptor patterns \cite{ref26}, organization of
adapted immune systems \cite{ref27}, amorphous silicon
\cite{ref28, ref29}, a wide class of disordered cellular materials
\cite{ref30}, dynamic random organizating systems
\cite{hexner2017noise, hexner2017enhanced, weijs2017mixing,
lei2019nonequilibrium, lei2019random}, electron density distributions \cite{Ge19, sakai2022quantum}, vortex distribution in superconductors \cite{Ru19, Sa19}, and even the distribution of primes on the number axis \cite{torquato2019hidden}. Many of these systems were thoroughly discussed in the seminal review article by Torquato (see Ref. \cite{To18a}).


The unique structural characteristics of DHU systems, i.e., the combination of both high-symmetry liquid-like structures on small-scales and crystal-like hidden order on large scales, endow them with many exotic and desirable physical properties, including optical properties \cite{ref31, ref32, ref33}, transport properties \cite{ref34}, mechanical properties \cite{ref35} as well as optimal multi-functionalities \cite{ref36}. One of the most striking discoveries is that certain carefully designed DHU materials are able to realize large, complete and isotropic photonic band gaps, blocking all directions and polarizations \cite{ref31, ref32, ref33}, which was considered not possible in traditional wisdom and has significant ramifications for electronic and phononic band gaps in disordered materials \cite{tang2022soft}. This discovery sets a new paradigm for engineered disorder in photonic metamaterials \cite{wu2017effective, zhang2019metasurface} and optical applications \cite{zhang2019experimental, zhang2021hyperuniform, zhang2022reconfigurable}. The designer DHU materials have also been successfully fabricated or synthesized using different techniques \cite{ref37, ref38}. In a recent review article \cite{yu2021engineered}, the engineered disordered in photonic metamaterials within the framework of hyperuniformity have been thoroughly discussed.




In light of the existing comprehensive reviews \cite{To18a} and \cite{yu2021engineered}, one may question the necessity for another review on hyperuniformity. There are several important motivations for this focused review: Very recently, there are many new discoveries of disordered hyperuniformity in solid state materials, including amorphous carbon nanotubes \cite{nanotube}, amorphous 2D silica \cite{Zh20}, amorphous graphene \cite{Ch21}, defected transition metal dichalcogenides \cite{PhysRevB.103.224102}, defected pentagonal 2D materials \cite{Zh21}, and medium/high-entropy alloys \cite{chen2021multihyperuniform}. It has been found the DHU states of these materials often possess a significantly lower energy than other disorder models, suggesting their superior stability. Moreover, DHU can lead to unique electronic and thermal transport properties, which resulted from mechanisms distinct from those that have been identified for their crystalline counterparts. For example, DHU states are found to enhance electronic transport in 2D amorphous silica \cite{Zh20}. On the other hand, DHU medium/high-entropy alloys realize Vegard's law \cite{vegard1921konstitution}, and possess enhanced electronic band gaps and superior thermal transport properties at low temperatures \cite{chen2021multihyperuniform}. These unique properties open up many promising potential device applications in optoelectronics and thermoelectrics. Our current review nicely complements the existing ones by focusing on these important new developments in solid state materials. In addition, our discussion of the concept of hyperuniformity is approached from an applied and ``materials'' perspective, while Ref. \cite{To18a} focuses more on the physics aspects and Ref. \cite{yu2021engineered} focuses more on the photonics aspects.


The rest of this focus review is organized as follows: In Sec. II, we discuss the definition of hyperuniformity and its generalizations in the context of solid state materials, and subsequently discuss different types of disordered hyperuniformity in these materials. In Sec. III, we discuss the effects of imperfections, including thermal fluctuations, defects, and stochastic displacements in solid state materials on hyperuniformity. Such knowledge can provide valuable guidance to identify potential candidates of novel disordered hyperuniform solid state materials. In Sec. IV, we focus on DHU medium/high-entropy alloys. In Sec. V, we thoroughly review the recent developments in DHU 2D materials. In Sec. VI, we discuss quasi-1D DHU material systems such as carbon nanotubes. In Sec. VII, we provide concluding remarks and outlook of the field, with an emphasis on novel DHU quantum materials discovery and design.

\section{Disorder and Hyperuniformity}
\label{definition}

\subsection{Definition of Hyperuniformity}

As briefly mentioned in Sec. I, the concept of hyperuniformity involves measuring density fluctuations in a material system on different length scales. Direct measurement of density fluctuations requires knowledge of atomic positions, which for example can be extracted from electron microscopy micrographs for 2D materials \cite{Zh20, Ch21}. It is generally much more challenging to extract atomic positions for three-dimensional bulk materials. However, the hyperuniformity properties of such materials can be assessed from scattering intensity data \cite{ref28, ref29}, as we will discuss in detail below.

\begin{figure*}[ht]
\begin{center}
$\begin{array}{c}\\
\includegraphics[width=0.475\textwidth]{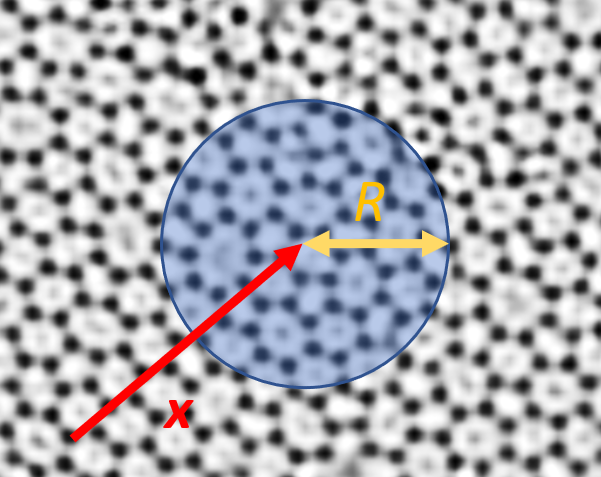}
\end{array}$
\end{center}
\caption{Illustration of the measurement of number density fluctuations in a material system using a spherical observation window $\Omega(R, {\bf x})$ with radius $R$ and centered at ${\bf x}$. Image courtesy of Pinshane Huang.} \label{fig2}
\end{figure*}

Without loss of generality, consider a transmission electron microscopy (TEM) image of a 2D amorphous silica, where the Si atomic columns appear as dark dots (see Fig. \ref{fig2}). The positions of Si atoms, represented as the centers of the dark dots, can be readily extracted using standard imaging processing \cite{Zh20}. This allows one to quantify the local Si atom density $n({\bf x})$ via
\begin{equation}
n({\bf x}) = \sum_i^N \delta({\bf x} -{\bf x}_i),
\end{equation}
where $N$ is the total number of atoms, $\{{\bf x}_i\}$ ($i=1, \ldots, N$) is the positions of the atomic centers, and $\delta({\bf x})$ is the Dirac delta function. In other words, $n({\bf x})$ is a collection of $N$ $\delta$-functions placed at the atom centers, which ``pick up'' the atoms.

Now we place spherical observation window $\Omega(R,{\bf x})$ of radius $R$ at location ${\bf x}$ in the system. The window only registers the atoms within itself, which is mathematically represented by the window indicator function:
\begin{align}
  \omega(R, {\bf x}, {\bf y})& = \left\{\begin{array}{l l}
    1,      & |{\bf y} - {\bf x}|\le R\\
    0,      & |{\bf y} - {\bf x}|> R.
  \end{array}\right.
  \label{eq:window}
\end{align}
The number of atomic centers falling into the window is then given by
\begin{equation}
N(R, {\bf x}) = \int_V {\omega(R, {\bf x}, {\bf y})}\cdot n({\bf y}) d{\bf y}.
\end{equation}
Therefore, $N(R, {\bf x})$ is random variable that depends on the observation window location ${\bf x}$ as the window is randomly placed in the system. This allows one to compute the mean number $\langle N(R) \rangle$ by averaging $N(R, {\bf x})$ over the window locations, where $\langle \cdot \rangle$ denotes the volume average. We note that the volume average is equivalent to ensemble average for ergodic systems \cite{To03}, which is the focus here.

The local number variance $\sigma^2_N(R)$ is then defined as
\begin{equation}
\sigma^2_N(R) = \langle N^2(R) \rangle - \langle N(R) \rangle^2.
\end{equation}
It has been shown that $\sigma^2_N(R)$ possesses the following asymptotic expansion in the large-$R$ limit:
\begin{equation}
\sigma^2_N(R) \sim A\cdot R^{d} + B\cdot R^{d-1} + \ell(R^{d-1}),
\label{eq_sigma2}
\end{equation}
where $A\ge 0$, $B>0$ and $\ell(x)$ indicates terms with orders lower than $x$. In Ref.\cite{To03}, it was shown that
\begin{equation}
A = \lim_{|{\bf k}|\rightarrow 0}S({\bf k}),
\end{equation}
where $S({\bf k})$ is the {\it static structure factor} associated with the atom centers, which can be obtained via scattering experiments (by excluding the forward scattering intensity). Specifically, it is defined as
\begin{equation}
S({\bf k}) = \frac{1}{N}\langle { \sum_{i\neq j = 1}^N \exp[-i {\bf k}\cdot({\bf x}_i - {\bf x}_j)]}\rangle .
\label{eq_Sk}
\end{equation}
Another relevant statistical descriptor for the atomic distribution is the {\it pair correlation function} $g_2({\bf r})$, which is defined via
\begin{equation}
\rho g_2({\bf r}) = \frac{1}{N}\langle { \sum_{i\neq j = 1}^N \delta({\bf r} + {\bf x}_i - {\bf x}_j)}\rangle,
\label{eq_g2}
\end{equation}
where $\rho = N/V$ is the atomic number density. Intuitively, $g_2$ measures the average local number density fluctuations with respect to the global number density $\rho$. Combining Eq.~(\ref{eq_Sk}) and Eq.~(\ref{eq_g2}) leads to the following relation:
\begin{equation}
S({\bf k}) = 1 + \rho\Tilde{h}({\bf k}),
\label{eq_Sk_h}
\end{equation}
where $\Tilde{h}({\bf k})$ is the Fourier transform of the total correlation function $h({\bf r}) \equiv g_2({\bf r}) - 1$. Eq. (\ref{eq_Sk_h}) indicates that $S({\bf k})$ and $g_2({\bf r})$ encode the same amount of information on pair correlations in the material systems.


A $d$-dimensional material is hyperuniform if its atomic number variance $\sigma_N^2(R)$ grows in the large-$R$ limit slower than $R^d$, i.e.,
\begin{equation}
\lim_{R\rightarrow\infty} \frac{\sigma_N^2(R)}{R^d} = 0.
\label{eq_hyper}
\end{equation}
This behavior is to be contrasted with those of typical disordered systems, such as Poisson point patterns, gases and liquids, where the number variance scales as $R^d$, i.e., like the volume of the observation window. It follows immediately from Eq.~(\ref{eq_sigma2}) and Eq.~(\ref{eq_hyper}) that the necessary and sufficient condition of hyperuniformity is that the zero-wavenumber limit of the static structure factor vanishes, i.e.,
\begin{equation}
A = \lim_{|{\bf k}|\rightarrow 0}S({\bf k}) = 0.
\label{eq_Sk0}
\end{equation}
In many applications, condition (\ref{eq_Sk0}) is easier to check compared to the direct computation of $\sigma_N^2$ and assessing the large-$R$ limit.

Hyperuniform systems can be further classified according to the small wavenumber (small $|k|$) behaviors of the structure factor $S({\bf k})$, which are equivalent to the asymptotic large $R$ behaviors of the number variance $\sigma_N^2(R)$. In particular, consider systems that are characterized by a structure factor with a radial power-law form in the
vicinity of the origin, i.e., $S({\bf k}) \sim |{\bf k}|^\alpha$ for $|{\bf k}|\to 0$. For hyperuniform point configurations, the exponent $\alpha$ is positive ($\alpha>0$)
and its value  determines three different large-$R$ scaling behaviors  of the
number variance \cite{To03, Za09, To18a}:
\begin{align}
  \sigma_{N}^2(R)&\sim\left\{\begin{array}{l l}
    R^{d-1},      &\alpha>1 \text{ (class I)}\\
    R^{d-1}\ln R, &\alpha=1 \text{ (class II)}\\
    R^{d-\alpha}, &0<\alpha<1 \text{ (class III).}
  \end{array}\right.
  \label{eq:sigma-scaling}
\end{align}
These scalings of $\sigma_{N}^2(R)$ define three classes of
hyperuniformity~\cite{To18a}, with classes I and III describing the
strongest and weakest forms of hyperuniformity, respectively.
States of matter that belong to class I include all perfect
crystals \cite{To03,Za09}, many perfect quasicrystals \cite{Za09,Li17,Og17}, and ``randomly" perturbed crystal structures  \cite{Ga04b,Ga08,Ki18a},
classical disordered ground states of matter \cite{To03,To15} as well as
systems out of equilibrium \cite{Zh16a,Le19a}. Class II hyperuniform systems include some quasicrystals \cite{Og17}, and many disordered classical \cite{Do05,Za11a,Ji11} and quantum \cite{Fe56,Re67,To08} states of matter. Examples of class III hyperuniform systems include random organization models \cite{He15} and perfect glasses \cite{Zh16a}.

\subsection{Different Types of Disorder in Solid-State Materials and Generalizations of Hyperuniformity}


\begin{figure*}[ht]
\begin{center}
$\begin{array}{c}\\
\includegraphics[width=0.95\textwidth]{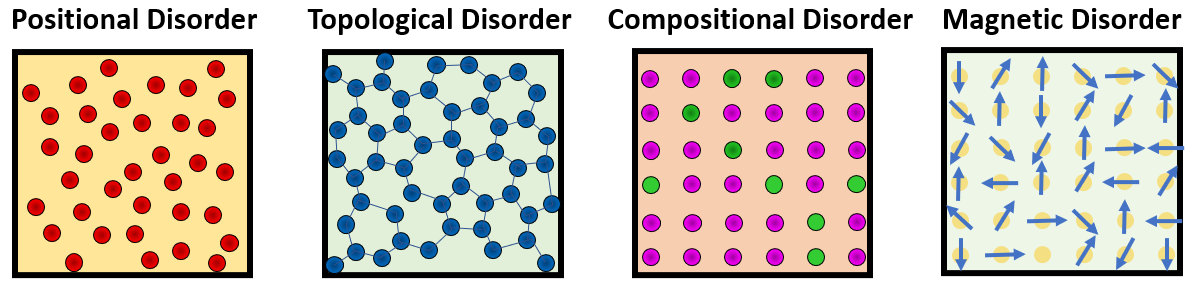}
\end{array}$
\end{center}
\caption{Illustration of different types of disorder commonly seen in amorphous solid state materials.} \label{fig3}
\end{figure*}

The notion of hyperuniformity was originally introduced for point configurations that possess {\it positional disorder} \cite{To03}, see Fig. \ref{fig3}(a). This type of disorder is also very commonly seen in amorphous solid state materials, such as metallic glasses, with respect to the distribution of the atomic centers. Another type of disorder commonly seen in covalent materials is {\it topological disorder} (see Fig. \ref{fig3}(b)), such as in amorphous silicon \cite{ref28, ref29} and 2D materials \cite{Zh20, Ch21}. In these materials, local topological defects that preserve the local chemical order (e.g., the coordination shell) are accumulated, eventually leading to topologically disordered networks, consisting of atomic loops of different sizes (e.g., 5-fold, 6-fold and 7-fold rings). In these systems, the topological disorder also induces positional disorder in atomic center distributions, which can treated using the same framework for quantifying number variance discussed in Sec. II.A.

{\it Compositional disorder} is typically found in solid solution states of alloys (see Fig. \ref{fig3}(c)), in which atoms of different alloying elements occupy the sites of certain crystal lattices (such as face-centered cubic and body-centered cubic lattices) \cite{porter2009phase}. Different atomic species can favor either like- or unlike-neighbors, as a result of the competition between enthalpic and entropic effects, leading to local composition fluctuations associated with different atomic species. Although the compositional disorder can be characterized by analyzing the atomic number variance for each component as well, it is also convenient to introduce scalar fields to quantify, e.g., the local molar fractions of different atomic species.

{\it Magnetic disorder} or {\it spin disorder} is a type of commonly seen disorder in magnetic materials and randomly dilute magnetic alloys \cite{mezard1987spin, mezard1984nature} (see Fig. \ref{fig3}(d)). Each atom in a magnet possesses an intrinsic magnetic moment or spin, the orientation of which can be either binary (i.e., the Ising model \cite{peierls1936ising, aharoni2000introduction}) or continuous (i.e., the Heisenberg model \cite{heisenberg1985theorie, aharoni2000introduction}). If the Hamiltonian governing the interactions between the spins possesses simple forms (e.g., only including nearest neighbor interactions), the system can easily achieve ground state with long-range spin orientation order such as the ferromagnetic or anti-ferromagnetic state. On the other hand, if the Hamiltonian includes non-trivial long-range spin-spin interactions, the system typically ends up in a metastable state with frozen-in spin disorder, i.e., a spin glass. Spin glasses can also be produced by alloying magnetic elements with non-magnetic elements. In this case, the magnetic disorder is induced by the compositional disorder in the alloy \cite{binder2011glassy}. It is also convenient to describe the magnetic system as a random scalar field, where the scalar values indicate the local spin orientation states. In this case, the hyperuniform framework based on point configuration cannot be applied and a proper generalization of the notation to continuous random field is required.

The hyperuniformity concept was first generalized by Torquato and co-workers to binary random fields \cite{Za09} and subsequently to continuous scalar fields \cite{To16a}. Consider a statistically homogeneous random scalar field $F({\bf x})$ in $d$-dimensional space that is real-valued. Similar to the case of point configurations, we employ a spherical observation window with radius $R$ and volume $v_1(R) \sim R^d$. In this case, the quantity of interest is the {\it local field intensity}, defined as the integration of field within the observation window over the window volume, which fluctuates as the window randomly moves in the system. The associated variance $\sigma_F^2(R)$ is given by
\begin{equation}
\sigma_F^2(R) = \frac{1}{v_1(R)}\int_{\mathbb{R}^d} \psi({\bf r})\alpha(r; R)d{\bf r},
\end{equation}
where $\alpha(r; R)$ is the scaled intersection volume, i.e., the intersection volume of two spherical windows of radius $R$ whose centers are separated by a distance $r$, divided by the volume $v_1(R)$ of the windows. A hyperuniform scalar field is one whose
$\sigma_F^2(R)$ decreases more rapidly than $R^{-d}$ for large $R$, i.e.,
\begin{equation}
\lim_{R\rightarrow\infty}\sigma_F^2(R) \cdot R^d = 0.
\end{equation}

Equivalently, the hyperuniformity of scalar field can be defined via its autocovariance function, i.e.,
\begin{equation}
\label{eq_psi}
\psi({\bf r}) = \langle{(F({\bf x}_1)-\langle{F({\bf
x}_1)}\rangle)(F({\bf x}_2)-\langle{F({\bf x}_2)}\rangle)}\rangle,
\end{equation}
where ${\bf r} = {\bf x}_2-{\bf x}_1$. The associated spectral density
function $\hat{\psi}({\bf k})$ is given by
\begin{equation}
\hat{\psi}({\bf k}) = \int_{\mathbb{R}^d} \psi({\bf r})e^{-i{\bf
k}\cdot{\bf r}}d{\bf r},
\end{equation}
which is the Fourier transform of $\psi(r)$. The hyperuniform
condition is then given by
\begin{equation}
\label{eq_hyper2} \lim_{|{\bf k}|\rightarrow 0}\hat{\psi}({\bf k})
= 0,
\end{equation}
which implies that
\begin{equation}
\int_{\mathbb{R}^d} \psi({\bf r})d{\bf r} = 0.
\end{equation}
In the case of hyperuniform scalar field \cite{To16a},  there are three different scaling regimes (classes) that describe the associated large-$R$ behaviors of the local field intensity variance when the spectral density goes to zero as a power-law scaling  ${\tilde \psi}({\bf k})\sim |{\bf k}|^\alpha$ as $|\bf k|$ tends to zero:
\begin{align}
\sigma^2_{F}(R) \sim
\begin{cases}
R^{-(d+1)}, \quad\quad\quad \alpha >1 \qquad &\text{(Class I)}\\
R^{-(d+1)} \ln R, \quad \alpha = 1 \qquad &\text{(Class II)}\\
R^{-(d+\alpha)}, \quad 0 < \alpha < 1\qquad  &\text{(Class III).}
\end{cases}
\label{eq:classes}
\end{align}
Classes I and III are the strongest and weakest forms of hyperuniformity, respectively.

The above generalized framework for hyperuniform random fields can be readily employed to quantify the aforementioned compositional disorder and spin disorder in solid state materials. In addition, this framework also allows one to analyze other important quantities in amorphous materials, such electron density distributions \cite{Ge19, sakai2022quantum}. An interesting case is when the electrons are sufficiently localized in the vicinity of the atomic cores. In this case, the overall electron density distribution $F({\bf x})$
can be considered as linear superposition of the electron density distribution
$K({\bf x}, {\bf r}_i)$ around the atomic core
$i$ centered at ${\bf r}_i$, i.e.,
\begin{equation}
\label{eq_field} F({\bf x}) = \sum_i K({\bf x}, {\bf r}_i).
\end{equation}
It has been shown in Ref. \cite{To16a} that the spectral density
for a scalar field given by Eq. (\ref{eq_field}) can be written as
\begin{equation}
\label{eq_spectral} \hat{\psi}({\bf k}) = \rho \hat{K}^2({\bf
k})S({\bf k}),
\end{equation}
where $\rho$ is the number density of the atoms, $\hat{K}({\bf
k})$ is the Fourier transform of the local electron density function
$K({\bf x})$, and $S({\bf k})$ is the structure factor associated
with the atom distributions. It follows from this analysis that the localized electrons possess a hyperuniform density distribution in a material where the atoms are hyperuniformly distributed, which has been demonstrated in an analogous classical system \cite{jiao2021hyperuniformity}.

\section{Effects of Imperfections in Solid-State Materials on Hyperuniformity}

Real materials inevitably contain structural defects and are subject to thermal fluctuations, which could mask the the hyperuniformity nature of the underlying materials. Understanding the effects of such defects, thermal fluctuations and other perturbations could therefore provide valuable guidance for the identification and discovery of new disordered hyperuniform solid state materials. We will also discuss some general conditions that have been identified under which hyperuniformity could arise or be preserved. An example is the transformation of perfect crystalline state in 2D hexagonal materials by introducing topological defects, which has been shown to preserve the hyperuniformity property of the crystalline state \cite{Ch21}.

According to the well-known fluctuation–compressibility theorem \cite{Ha86}, the isothermal compressibility of equilibrium single-component many-particle ensembles $\kappa_T$ at temperature $T$ and number density $\rho$ is linked to the structure factor $S(k)$ through the following relationship:
\begin{equation}
    \rho k_B T \kappa_T = S({\bf k}={\bf 0}),
\end{equation}
where $k_B$ is the Boltzmann's factor. Consequently, thermal fluctuations at any finite temperature will destroy hyperuniformity in any compressible equilibrium systems (i.e., $\kappa_T \neq 0$), regardless of the degree of order in the system \cite{To18a}. Moreover, strong theoretical arguments \cite{To18a} indicate that the large-distance asymptotic behavior of the direct correlation function $c({\bf r})$ of disordered phases is exactly proportional to the potential $v({\bf r})$ and the volume integral of $c({\bf r})$ does not exist for hyperuniform systems. Therefore, effective long-ranged interactions are necessary to achieve incompressibility and drive an equilibrium many-particle system to a hyperuniform state. However, this condition is not necessary to achieve hyperuniformity in systems out of equilibrium \cite{To18a}. Here, the direct correlation function $c({\bf r})$ is linked to the total correlation function $h({\bf r})$ defined above via an integral equation \cite{Or14}:
\begin{equation}
    h({\bf r}) = c({\bf r}) + \rho c({\bf r}) \otimes h({\bf r}),
\end{equation}
which upon Fourier transformation evolves into
\begin{equation}
    \tilde{h} ({\bf k}) = \frac{\tilde{c} ({\bf k})}{1-\rho\tilde{c} ({\bf k})}.
\end{equation}

In the context of solid state materials, disorder often comes in the form of a wide spectrum of defects and displacements introduced into crystals, and here we discuss their effect on the preservation and generation of hyperuniformity. For point defects such as vacancies and substitutions, it is known that they will destroy hyperuniformity in proportion to the concentration $p$ of vacancy or substitutions when they are introduced randomly into the crystals, i.e., there is no spatial correlation between vacancies or substitutions \cite{Ki18}:
\begin{equation}
    S({\bf k}) = p + (1-p) S_0({\bf k}),
\end{equation}
where $S_0({\bf k})$ is the structure factor of the original cyrstalline state.
On the other hand, randomly introduced interstitials will destroy hyperuniformity proportional to $p/(1+p)$ \cite{Ki18},
\begin{equation}
    S(k) = \frac{p}{1+p} + \frac{1}{1+p} S_0({\bf k}),
\end{equation}
where $p$ is the ratio of the number of interstitials over the number of particles originally in the crystals. However, if vacancies, substitutions, or interstitials are introduced in a hyperuniform manner, then the resulting patterns will also preserve the hyperuniformity of the defect-free crystals \cite{Ki18}.

``Uncorrelated'' stochastic displacements in perfect lattices, i.e., the displacements of distinct particles are uncorrelated and the joint probability density function can be reduced into a product of two singlet probability densities \cite{Ki18}, have also been employed to model perturbed crystalline states of solid materials. It has been shown that they can never destroy the hyperuniformity but it can be degraded such that the perturbed lattices fall into class-III hyperuniform systems \cite{Ga04a, Ki18, Kl20}. On the other hand, correlated displacements (e.g., thermal excitation in compressible crystals or mechanical vibrations) can destroy hyperuniformity \cite{Ki18, PhysRevB.103.224102}. However, under certain special conditions correlated displacements can also preserve hyperuniformity \cite{chen2021topological}. For example, it has been shown that hyperuniformity is preserved when the following three conditions are met: (i) sources of the displacements are randomly introduced and sparse; (ii) the displacements are sufficiently localized (i.e., the volume integrals of the displacements and squared displacements caused by individual defect are finite); (iii) the displacement-displacement correlation matrix of the system is diagonalized and isotropic.

It is also noteworthy that a variety of hyperuniformity-generating operations that can convert nonhyperuniform point patterns into hyperuniform materials have also been reported. For example, Klatt et al. \cite{ref30} demonstrated that one can convert a wide spectrum of nonhyperuniform and hyperuniform point patterns into effectively stealthy hyperuniform patterns by applying a sequence of correlated displacements that leads to the generation of centroidal voronoi tessellations. Kim and Torquato \cite{Ki19a} have discovered that by decorating each point in a point pattern with spheres of different sizes so that the local packing fraction is identical across the sample, one can convert a variety of point patterns into perfectly hyperuniform materials. This operation shared some similarities with the ``equal-volume tessellation'' operation \cite{Ga08} that can be used to obtain hyperuniform point patterns.

\section{Disordered Hyperuniform Medium/High-Entropy Alloys}




High-entropy alloys (HEAs) are those composed of mixtures of equal or relatively large proportions of five or more elements. Alloys containing three elements are usually referred to as medium-entropy alloys (MEAs) \cite{ZHANG20141}. These MEAs and HEAs are distinct from traditional metallic alloys which contain one or two major components with smaller amounts of other elements in that they are stabilized by entropy effects as solid solution phases. Studies have shown that MEAs and HEAs can possess superior mechanical performance and resistance to corrosion compared to traditional alloys \cite{MIRACLE2017448}, and can have unusual thermal and electronic transport properties \cite{mu_pei_liu_stocks_2018}, opening up both new structural and functional applications.

Detailed atomic structures of medium/high-entropy alloys largely remain a mystery, due to challenges in direct imaging of bulk atomic packing arrangements in three dimensions (3D). Over the past three decades, the random mixture model and special quasirandom structure (SQS) \cite{PhysRevLett.65.353, Wa13} have been widely used for HEA/HEA structures, assuming that atoms of different types in the alloys are randomly distributed on the sites of an underlying crystalline lattice. Recently, a number of studies (e.g., on chromium-cobalt-nickel (CrCoNi) MEAs and silicon-germanium-tin (SiGeSn) MEAs \cite{jin2022coexistence}) have discovered significant short-range order (SRO). These SROs are often manifested as suppressed clustering of atoms of the same type, which are shown to lead to energetically favorable states of the alloys and are not captured by random mixture and SQS models \cite{ding2018tunable,zhang2020short, walsh2021magnetically, Be21}. A direct consequence of the existence of such short-range order is the suppression of composition (or density) fluctuations. Since the atoms prefer certain neighbor arrangements, it can be expected that as an observation window moves randomly within the system, the number of atoms of a particular element that fall within the window would not fluctuate significantly, especially for large window sizes, which is consistent with the concept of hyperuniformity.

\subsection{Multi-hyperuniform Structural Model}

\begin{figure}[ht!]
\begin{center}
$\begin{array}{c}\\
\includegraphics[width=0.95\textwidth]{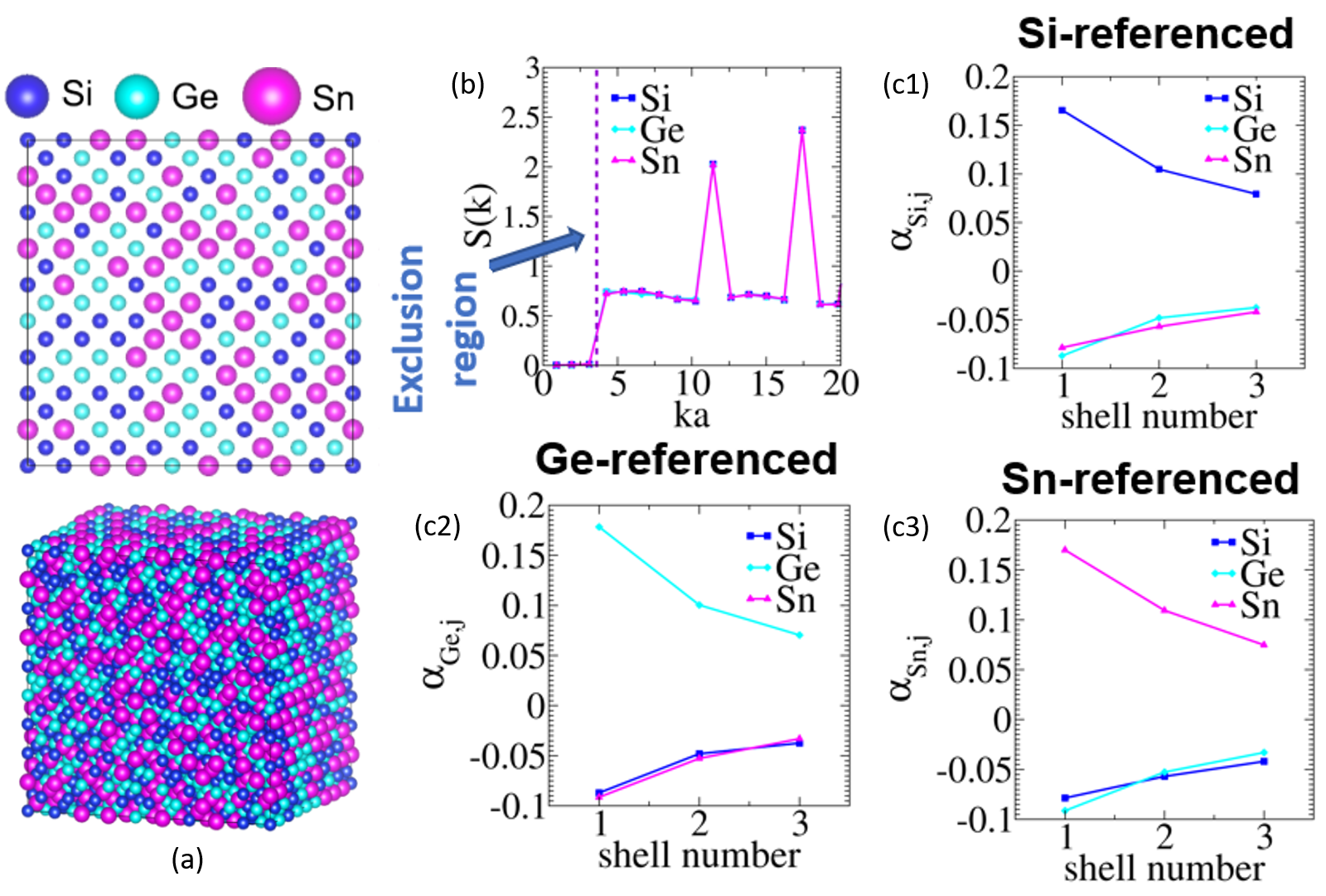}
\end{array}$
\end{center}
\caption{(a) Three-dimensional visualizations (bottom panel) and their two-dimensional slices (top panel) along the (001) plane of representative multihyperuniform (with an exclusion region of $K_0a=3.6$ for targeted structure factor, where $a$ is the lattice constant of a diamond cubic unit) SiGeSn alloys on a 9~$\times$~9~$\times$~9 diamond lattice. (b) Structure factors and (c) Warren-Cowley short-range order parameters $\alpha_{ij}^{\nu}$ of multihyperuniform alloy model.} \label{fig_MEA1}
\end{figure}

In Ref. \cite{chen2021multihyperuniform}, it was shown that SROs can lead to significant suppression of large-scale composition/density fluctuations, using SiGeSn as an example. In particular, Monte Carlo (MC) quenching simulations using Stillinger-Weber empirical potential calibrated for SiGeSn alloys \cite{tomita2018development}: Starting from a random solution state, a random selected pair of atoms of different types are swapped and the resulting energy change is calculated. Only swaps leading to decreasing energy are accepted, which amounts to a stochastic steepest descant in the energy landscape. Quenching simulations were used because the solid solution phases of the alloy, even with significant SROs, are not thermodynamically stable at low temperatures \cite{jin2022coexistence}. The quenched alloys possess significant SROs, quantified via the Warren-Cowley SRO parameter $\alpha_{ij}^{\nu}$, which is defined as
\begin{equation}
\label{eq_4} \alpha_{ij}^{\nu} = 1 - p_{ij}^{\nu}/c_j,
\end{equation}
where $p_{ij}^{\nu}$ is the probability of finding atomic species $j$ around an atom of type $i$ in the $\nu$-th neighboring shell. The systems also exhibit very strong (although not complete) suppression of large-scale composition fluctuations for all elements, a key feature of a {\it multihyperuniform} system. It is intuitive that the existence of SROs implies suppression of large scale fluctuations (and vice versa), as SROs indicate the elements are more homogeneously spread and thus, local clustering of same elements leading to large fluctuations is very rare.

Based by these observations, a multi-hyperuniform structural model for MEAs/HEAs with short-range orders were proposed in Ref. \cite{chen2021multihyperuniform}. In particular, a highly efficient generic Fourier-space construction technique was devised to generate large realizations of multihyperuniform MEAs (or HEAs): A fictitious ``energy'' $E$ of the system was defined as the squared differences between the target and constructed structure factors associated with different types of elements, i.e.,
\begin{equation}
\label{eq_E1} E = \sum_{i}\sum_{k} [S_i(k)-S_{i,0}(k)]^2,
\end{equation}
where $S_{i,0}(k)$ and $S_i(k)$ respectively are the angular-averaged structure factor associated with type-$i$ atoms in the target and constructed structures. The simulated annealing procedure was employed to evolve the system under construction (by randomly swapping atoms of different types) towards one with targeted structure factors via a stochastic optimization of the ``energy'' $E$. To obtain multihyperuniform MEAs or HEAs, the target structure factors $S_{i,0}(k)$ was set to be
\begin{equation}
\label{eq_Sk2} S_{i,0}(k)=0 ~~~~ \textnormal{for} ~~ k < K_{i,0},
\end{equation}
where $K_{i,0}$ is the range of exclusion region in the Fourier space for the target structure factor associated with element $i$. Cubic simulation boxes with length $L$ and periodic boundary conditions were employed, for which the wavevector $\bf{k}$ only takes discrete values ${\bf k} = \frac{2\pi}{L}(n_x, n_y, n_z)$, where $n_x$, $n_y$, and $n_z$ are integers. To mitigate the statistical noise of $S({\bf k})$ at each individual ${\bf k}$, angular average of $S({\bf k})$ was performed to obtain $S(k)$ by binning ${\bf k}$ according to its magnitude $k\equiv |{\bf k}|$.

A representative example of equimolar SiGeSn alloys (i.e., $c_\mathrm{Si}=c_\mathrm{Ge}=c_\mathrm{Sn}=1/3$) with an underlying diamond lattice structure, obtained from the inverse construction (by setting $K_{i,0}=K_0 = 3.6/a$ for all atom species, where $a$ is the lattice constant of the diamond cubic unit cell) is shown in Fig. \ref{fig_MEA1}(a). The  structure factors associated with different elements are shown in Fig. \ref{fig_MEA1}(b), which decrease as $k$ decreases at small $k$ and essentially approaches zero as $k$ goes to zero, indicating hyperuniformity of the atomic distribution for all element types. The SRO parameters are shown in Fig. \ref{fig_MEA1}(c), which indicates the multihyperuniform structures possess substantial SRO, i.e., atoms of the same type are disfavored in the first three neighbor shells around any atom in the system. The emergent SRO in MEAs is induced by multihyperuniformity (i.e., no constraints were imposed to target the realization of SRO), suggesting that these two types of order may be coupled. This suggests in order to achieve suppression of composition fluctuations on large length scales for each species on a lattice, the system needs to suppress local clustering of the same type of atoms. This rationale is not unique to the SiGeSn alloy, and given the ubiquitous nature of SRO discovered previously in MEAs and HEAs \cite{ding2018tunable,zhang2020short, walsh2021magnetically, Be21}, strongly indicates the possibility of hidden hyperuniformity in other MEAs or HEAs with SROs.


\subsection{Physical Properties}

\begin{figure}[ht!]
\centering
\includegraphics[width=0.95\textwidth,keepaspectratio]{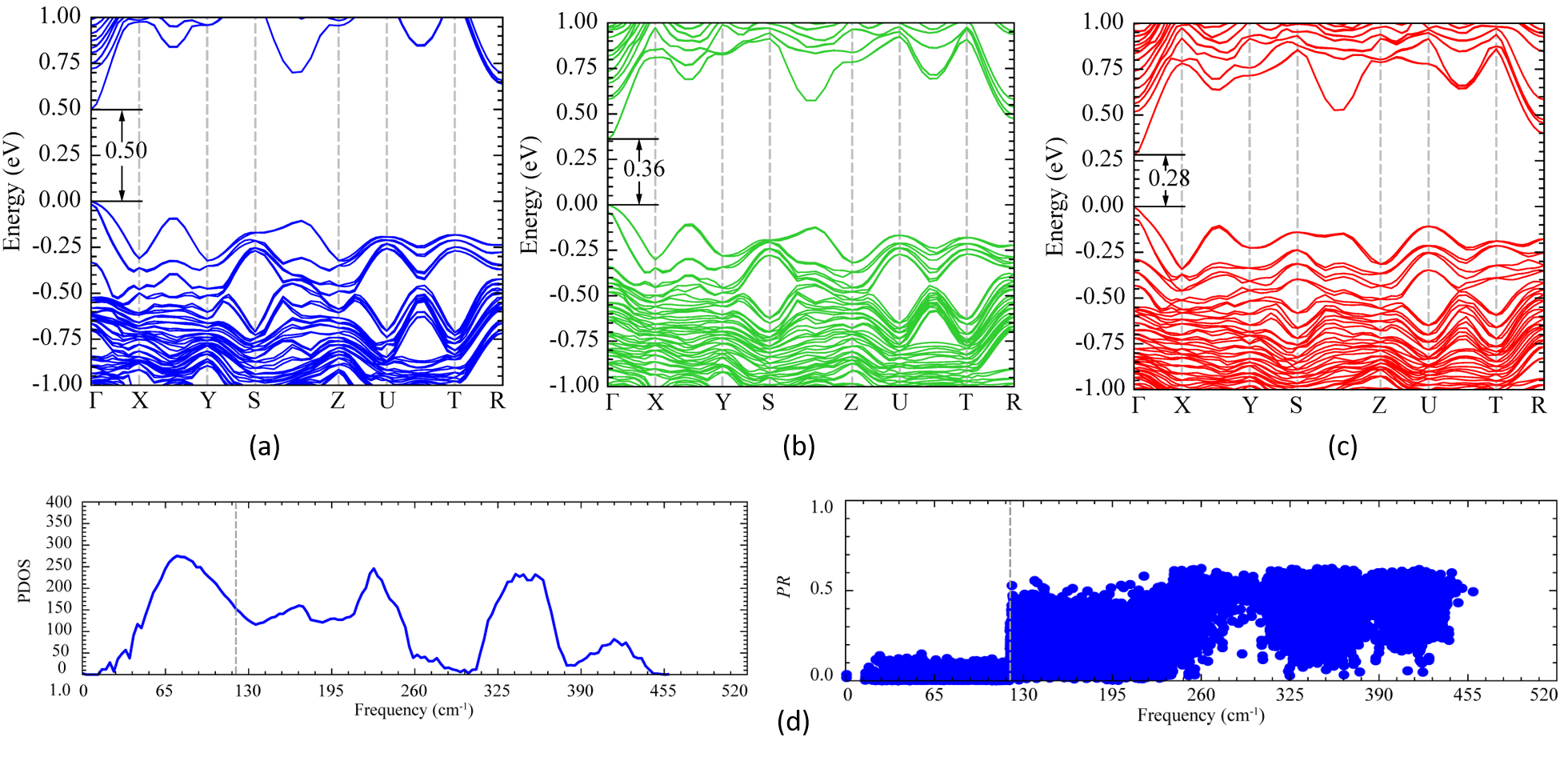}
\caption{Physical properties of multihyperuniform SiGeSn alloys: Band structures of the multihyperuniform (a), SQS (b), and random (c) models for SiGeSn MEAs. (d) Phonon density of states (PDOS) and partition ratio (PR) of SiGeSn MEA with a multihyperuniform structure.}
\label{fig_MEA2}
\end{figure}

\subsubsection{Stability and lattice distortions}

In Ref. \cite{chen2021multihyperuniform}, the authors calculated the formation energy of the multihyperuniform SiGeSn MEAs with respect to a linear combination of the energies of pure Si, Ge, and Sn to be 0.089 eV/atom using DFT calculations of systems with 216 total atoms. The positive values indicate the solution phases of the alloy are thermodynamically unstable to phase separation at 0 K, and thus are metastable states that only can be obtained by quenching \cite{jin2022coexistence}. It was also discovered that the multihyperuniform structures with suppressed composition fluctuations should lead to a smaller deviation from Vegard's law and smaller lattice distortion that is intrinsic to all HEA or MEA systems \cite{latticedistortion}, compared to other solution models. Specifically, the structures of different models were optimized using DFT simulations and the lattice constants of the optimized multihyperuniform, random, and SQS structures with $3\times3\times3$ supercells were determined to be 17.903, 17.883, and 17.883~\AA, respectively. According to Vegard's law \cite{vegard1921konstitution}, the lattice constant of an equimolar SiGeSn alloy is predicted to be 17.904~\AA, the weighted average of the lattice constants of pure Si (16.406~\AA), Ge (17.346~\AA), and Sn (19.960~\AA) crystals with $3\times3\times3$ supercells, which clearly indicates that the multihyperuniform system exhibits an minimal amount of deviation from Vegard's law and reflects the nearly ideal mixing of the three constituent elements in the multihyperuniform SiGeSn MEA. The lattice distortion was also computed using the vector dissimilarity metric \cite{zhuang2021sudoku}, i.e., the root mean squared displacement (RMSD) between the reference lattice (a diamond lattice with a lattice constant as predicted by Vegard's law) for the multihyperuniform, random, and SQS systems, which are 2.14, 3.21, and 3.31 \AA, respectively. Clearly, the multihyperuniform system possesses the smallest lattice distortion, which leads to lower energy of such system than the (quasi)random systems.


\subsubsection{Electronic bandgaps}

The effect of multihyperuniform long-range order (MHLRO) on the electronic structure of the SiGeSn MEA was investigated by calculating the band structure of the SiGeSn MEA with the multihyperuniform, random, and SQS structure. Figures \ref{fig_MEA2}(b), (c), and (d) reveal direct band gaps of 0.28, 0.36, and 0.50 eV at the $\Gamma$ point for the random, SQS, and multihyperuniform systems, respectively. For pure crystalline Si, Ge, and Sn, the band gaps are 1.29, 0.33, and 0.15 eV, respectively. In the random and SQS model for the SiGeSn alloy, the clustering of Ge and Sn atoms significantly reduces the band gap, while in the multihyperuniform model the presence of  leads to suppressed local clustering of atoms of the same type, resulting in a band gap that is nearly the average band gap (0.59 eV) of the three constituent elements. This is another manifestation that the MHLRO SiGeSn alloy approximately realizes Vegard's law.

\subsubsection{Low-temperature thermal conductivity}

The thermal conductivity of the multihyperuniform system at low temperatures is generally higher than that of the random solution models, which are 1.53 and 0.87 W/(m$\cdot$ K) at 10 K for the multihyperuniform and random systems, respectively. The improved thermal conductivity makes the multihyperuniform SiGeSn MEA a better candidate for electronic device applications such as micro-chips that require a stricter management in heat dissipation. The higher thermal conductivity of the multihyperuniform system can be attributed to the smaller lattice distortions of such system compared to the random mixture, since large lattice distortions would enhance the scattering of phonons and lead to low thermal conductivity \cite{Fa16}. However, as the temperature increases, the effect of phonon-phonon coupling becomes more prominent, which leads to similarly decreased thermal conductivities for multihyperuniform and random mixtures. For example, it is found that the computed thermal conductivities at 400 K for the multihyperuniform and random systems are 0.53 and 0.57 W/(m$\cdot$ K), respectively. Figure \ref{fig_MEA2}(e) shows the phonon density of states (PDOS) and partition ratio (PR, which for each vibrational mode $n$ is calculated as $PR_n = (\sum_ie_{i,n}^2)^2/N\sum_ie_{i,n}^4$, where $e_{i,n}$ denotes the phonon eigenvector of mode $n$ and $N$ refers to the total number of atoms \cite{lee2017molecular,guo2015approaching}) of the multihyperuniform mixture. Although the overall shape of the PDOS is similar to that of the random system \cite{WANG2020443}, a sudo gap around 280 cm$^{-1}$ (corresponding to $T \sim 400$ K) was observed, leading to diminishing PR modes in the vicinity of this frequency, and a reduction in the thermal conductivity.


\section{Disordered Hyperuniform 2D Materials}

\subsection{Modeling Amorphous 2D Materials via Hyperuniform Networks Containing Stone-Wales Topological Defects}



\begin{figure}[ht]
\includegraphics[width=0.475\textwidth,keepaspectratio]{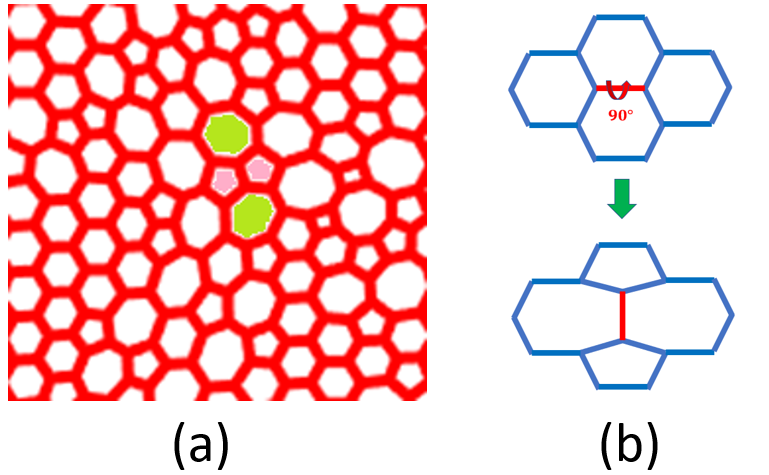}
\caption{(a) Structural model of amorphous 2D material obtained
by introducing SW defects in a perfect honeycomb network. (b)
Illustration of a SW defect, which changes the local network
topology and leads to a cluster of two pentagons and two
heptagons.} \label{fig_SW}
\end{figure}

2D materials, or single layers of van der Waals materials, are solid state materials a single unit cell thick, which can be structurally modeled as 2D networks. A commonly seen type of disorder in 2D materials results from topological defects, such as the Stone-Wales (SW) defects, which can be introduced in the system via, e.g., proton radiation (see Fig. \ref{fig_SW}) \cite{St86}. In the case of hexagonal 2D materials, the resulting structures contain ``flipped'' bonds that change the local topology of the
original honeycomb network, leading to paired clusters of two pentagons
and two heptagons. The Stone-Wales-like defects have been observed in a number of 2D
material systems \cite{Hu12, Hu13, Ed14, Zh15c, To20}, and their effects on the materials' physical properties have been numerically investigated \cite{Li11, PhysRevB.86.121408}.

\begin{figure}[ht]
\includegraphics[width=0.85\textwidth,keepaspectratio]{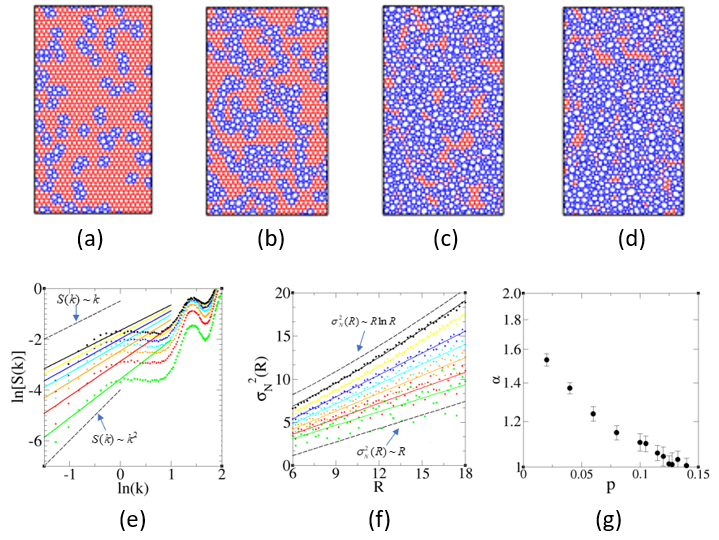}
\caption{Representative amorphous structural models at different defect fractions $p = 0.02$ (a), $p = 0.04$ (b), $p = 0.12$ (c), $p = 0.14$ (d), with the SW defect bonds highlighted in blue. Note that at small $p$, the defect bonds form small isolated clusters, while the defect bonds percolate and form large interconnected clusters at large $p$. (e) The static structure factor $S(k)$ of
these networks possesses the scaling $S(k) \sim k^\alpha$ for
small wave number $k$, where $1\le\alpha(p)\le2$ is
monotonically decreasing as the SW defect concentration $p$
increases. (f) The corresponding number variance $\sigma_N^2(R)$ possesses the scaling $\sigma_N^2(R) \sim R^{\gamma}$ with $\gamma<2$. (g) The scaling exponent $\alpha \in [1, 2]$ in $S(k)$ decreases as $p$ increases, reaches a value of 1.0 at $p \approx 0.12$, and then remains relatively flat as $p$ increases beyond 0.12. These results indicate that the Stone-Wales defects preserve hyperuniformity in amorphous 2D networks, and the existence of of hyperuniformity class transition from Class I to Class II at $p\sim 0.12$.}
\label{fig_SW2}
\end{figure}

The global structures of amorphous 2D materials resulted from these local defects were comprehensively and systematically investigated in Ref. \cite{Ch21}. Specifically, the structural models for amorphous 2D network structures were generated by randomly introducing SW defects (i.e., bond-ration induced topological transformations) into the perfect honeycomb lattice until a specific defect concentration $p$ is achieved, which was followed by structural relaxation to minimize a harmonic energy that regularizes the bond lengths and angles. It was found that all of the generated structures are hyperuniform, manifested as $\lim_{k\rightarrow0}S(k) = 0$ and the scaling $\sigma_N^2(R) \sim R^{\gamma}$ with $\gamma<2$ (see Figs. \ref{fig_SW2}). These results indicate that the SW transformation preserve hyperuniformity in the resulting amorphous network materials. This is consistent with the observation that SW defects are local perturbations, and thus, do not fundamentally change the nature of density fluctuations on large length scales compared to the original honeycomb lattice, which is hyperuniform \cite{chen2021topological}.

The static structure factor of the amorphous networks possesses the scaling $S(k) \sim k^\alpha$ for small wave number $k$ for all defect concentrations studied here, with the scaling exponent $\alpha \in [1, 2]$. $\alpha$ first decreases as $p$ increases, reaches 1.0 at $p \approx 0.12$, and then remains relatively flat as $p$ increases beyond 0.12 (see Fig. \ref{fig_SW2}). The initial decrease of $\alpha$ is due to the increasing disorder, which is saturated as the defects percolate at $p_c \approx 0.11$. This is also equivalently reflected in the scaling of the number variance $\sigma_N^2(R)$ from $\sim R$ to $R\ln(R)$ for large $R$ as $p$ increases. Since the generation of these amorphous network material models didn't assume any material specific details, they can be employed to model the structures of a variety of hexagonal 2D materials. For example, the amorphous network structure with $p=0.06$ matches previously synthesized experimental amorphous graphene sample \cite{To20} well in terms of various statistics, which we discuss below.

\subsection{Amorphous Graphene}


\begin{figure}[ht]
\includegraphics[width=0.95\textwidth,keepaspectratio]{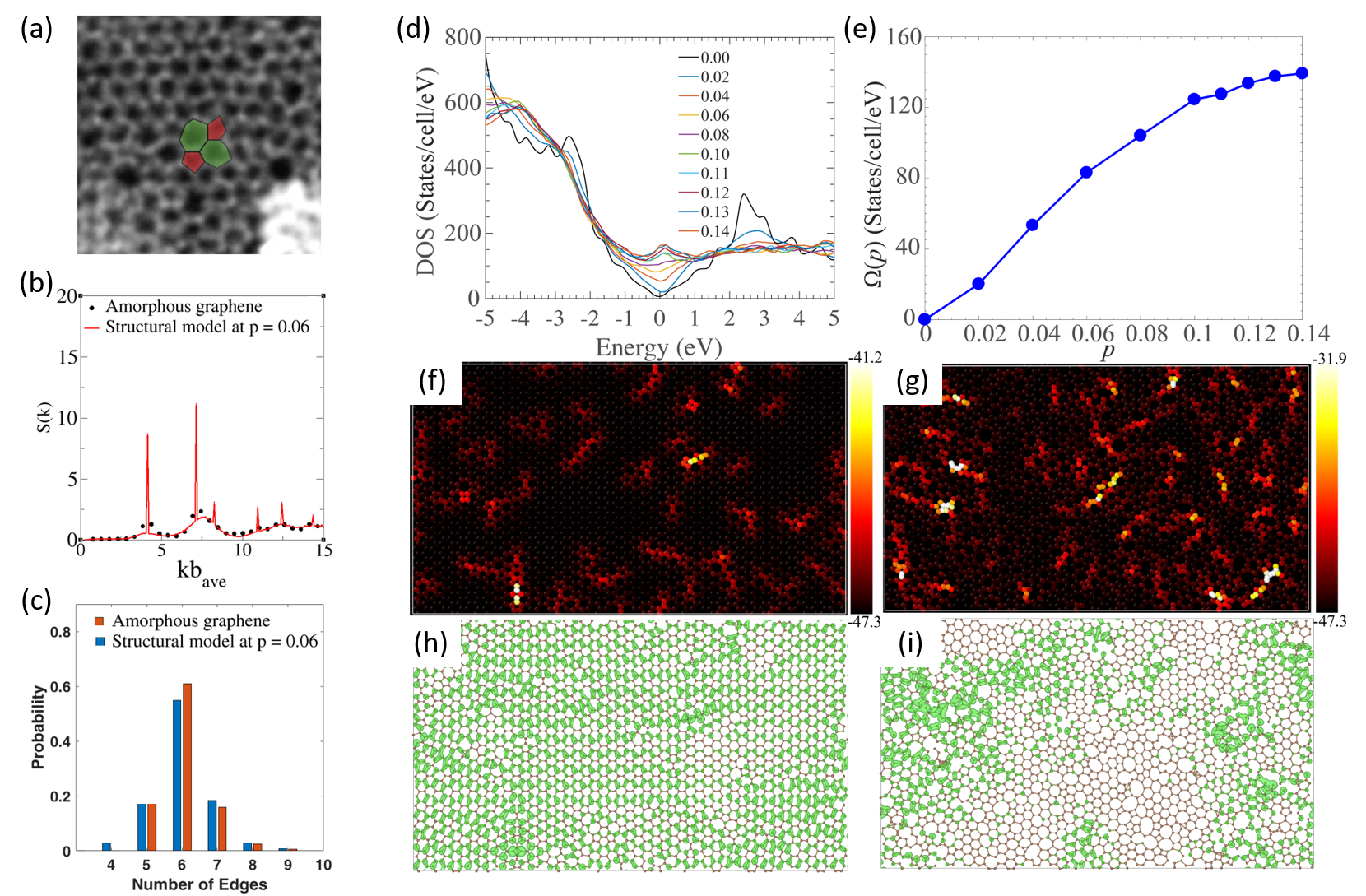}
\caption{Amorphous 2D materials containing the Stone-Wales
topological defects. (a) Scanning transmission electron microscopy (STEM) image of 2D amorphous graphene. Image courtesy of the Barbaros Oezyilmaz group. (b) Structure factor $S(k)$ of the amorphous graphene sample obtained in experiments and our structural model at $p=0.06$. (c) Polygonal shape distributions of the experimental graphene sample (red) and our structural model at $p=0.06$ (blue).  Density of states (DOS) of different concentrations (0$\leq$$p$$\leq$0.14) of Stone-Wales defects (d) in the energy window of -5 to 5 eV and (e) at the Fermi level denoted as $\Omega({p}$). Middle panels: Atomic energy distribution (in eV) of graphene with (f) $p$ = 0.02 and (g) $p$ = 0.12. Lower panels: Electron densities at the Fermi level of graphene with (h) $p$ = 0.02 and (i) $p$ = 0.12. The green surface represents isosurface of 1.0$\times10^{-5}$ atomic unit.} \label{fig_graphene}
\end{figure}


Recently, single-layer amorphous
graphene has been successfully synthesized \cite{To20}, which possesses a structure distinctly different from the random network model as revealed by detailed transmission electron microscopy characterization. In a wider context, amorphous
carbon-based systems have been extensively studied, including graphene
sheets \cite{Ha13b, Da16, Ma19}, cross-linked graphene network \cite{Hu17}, amorphous glassy carbon and carbon fibers \cite{Je71, Sa08}, to name a few. The generic structural networks discussed in Sec. IV.A can be converted into 2D amorphous material by decorating each vertex (as in the case of graphene) and/or the mid-point of each bond (as in the case of amorphous silica) in the network with an atom of a particular type. Examples of resulting 2D materials include graphene and graphene-like materials such as boron nitride (BN), molydenum disulphide (MoS$_2$), and silica (SiO$_2$).

The amorphous graphene model derived from our hyperuniform network structure at $p=0.06$ (class I hyperuniform) matches previously synthesized experimental amorphous graphene sample \cite{To20} well in terms of various statistics (see Fig. \ref{fig_graphene}(b) and (c)). Specially, the static structure factor $S(k)$ directly computed from the TEM image data reveals a degree of hyeruniformity in experimentally synthesized amorphous graphene, which agrees very well with the corresponding structure factor computed from the network model. Among other properties, the electronic structure properties of a wide spectrum of disordered hyperuniform graphene materials derived from our network models at different SW concentration values $p$ were investigated \cite{Ch21}.

Fig. \ref{fig_graphene}(d) shows the density of states (DOS) of the DHU graphene materials, which revels that the Dirac cone, i.e., the semi-metal nature of crystalline graphene, is destructed in in DHU graphene systems, which possess increasingly higher DOS at the Fermi level as $p$ increases. The DOS values $\Omega(p)$ at the Fermi level as a function of $p$ is shown in Fig. \ref{fig_graphene}(e), which exhibits monotonically increasing behavior as $p$ increases. The increased DOS at the Fermi level are also manifested in the energies and charge densities: The carbon atoms at the flipped C-C bonds and their adjacent regions exhibit higher energies. This is illustrated in Fig. \ref{fig_graphene}(f) and (g), respectively showing the atom-resolved total energies for two representative systems with $p$ = 0.02 and 0.12. These two amorphous graphene systems correspond class-I hyperuniformity ($p = 0.02$) and class-II hyperuniformity ($p = 0.12$), respectively. Figure \ref{fig_graphene}(h) and (i) respectively show the charge density at the Fermi level for these two systems. It can be seen that the electrons in class-I DHU graphene spread out in the entire system, while the electrons in class-II DHU graphene are localized in separate islands. These patches are similar to the localization regions found by Tuan et al. and shown to degrade the electrical transport of graphene \cite{Va12}.


These results indicate that increasing $p$ in the DHU graphene materials significantly populates the number of electron
states $\Omega(p)$ at the Fermi level, which is a result of the increasing number of high-energy states induced by the topological defects, and the Fermi-level charge densities indicate different electronic transport mechanisms associated with different classes of disorder, from patch-spreading to highly localized states. Such knowledge strengthens our fundamental understanding of change of behaviors in density fluctuations as defects are gradually introduced into ordered structures, and may have important implications for novel applications in photovoltaics, semiconductors, electrodes, batteries, water purification and multi-functional composites \cite{Bh15, Mi14, Xu13}. The degradation of mechanical properties in DHU graphene as $p$ increase was also investigated in Ref. \cite{Ch21}.


\subsection{Amorphous Silica}



\begin{figure}[htp]
\includegraphics[width=0.985\textwidth,keepaspectratio]{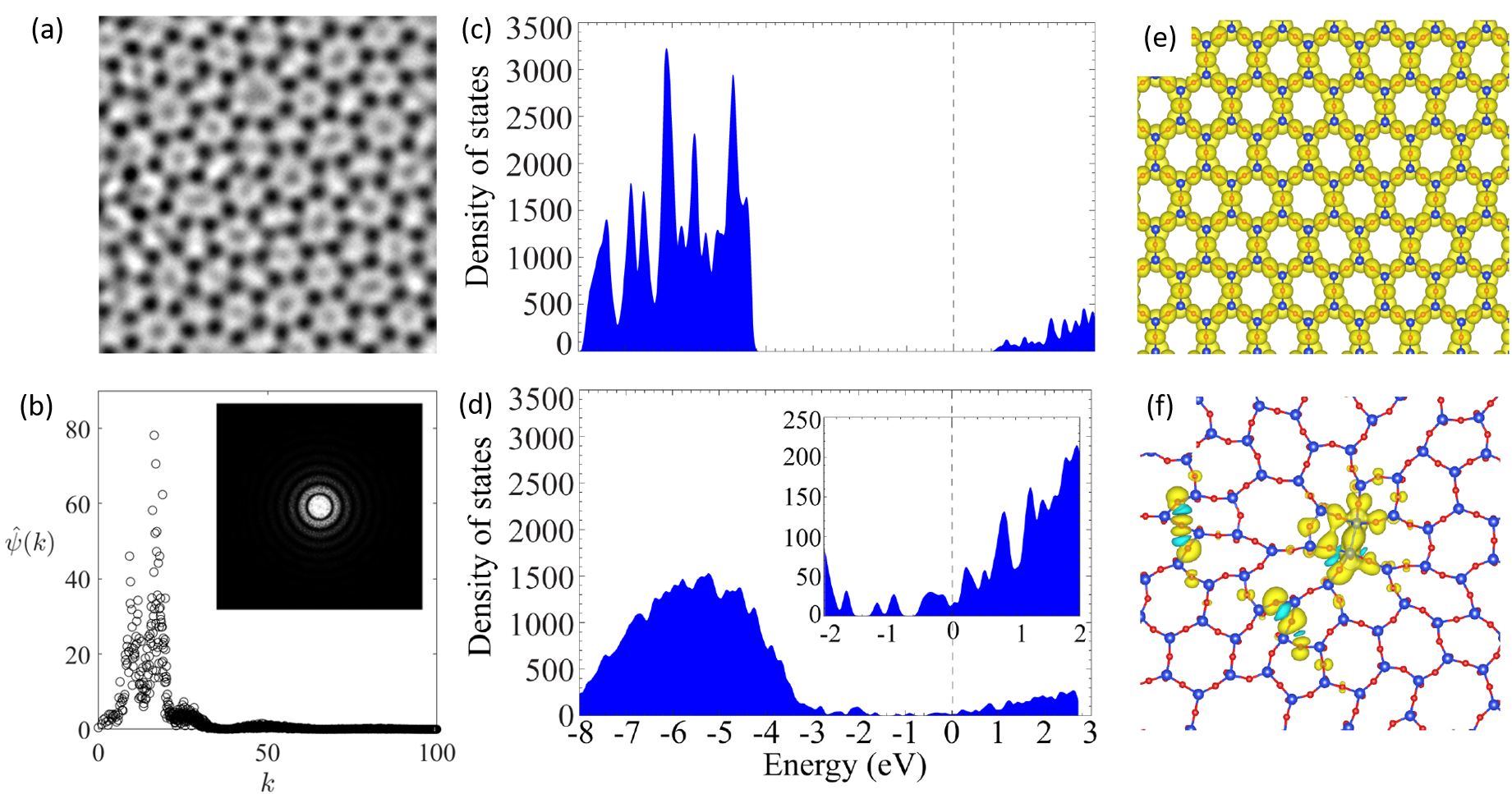}
\caption{Hyperuniformity in 2D amorphous silica. (a) TEM image of
2D amorphous silica. Reproduced from Ref. \cite{ref40}. (b) Angle-averaged spectral
density $\psi({\bf k})$ associated with the TEM micrograph and the
simulated amorphous silica network. Insets show the full spectral
density. Density of states of (c) the supercells of 2D crystalline
and (d) hyperuniform SiO$_2$ calculated using density functional
theory. The inset shows a zoom-in view of the DOS near the Fermi
level. Electron densities (e) at the HOMO level of the
crystalline structure and (f) at the Fermi level of the DHU
structure. The isosurface values is 0.5 $\times 10^{-8}$
$e$/Bohr$^3$. } \label{fig_slica}
\end{figure}

Recently, the discovery of hyperuniformity in
amorphous 2D silica (conventionally modeled as ``continuous random
networks'' \cite{ref39}), based on the analysis of aberration
corrected transmission electron microscopy (TEM) images of the
material was reported \cite{Zh20}, see Fig.\ref{fig_slica}. The amorphous silica
samples were fabricated using chemical vapor deposition (CVD) and
the procedure for obtaining the imaging data set was reported in
detail in Ref. \cite{ref40}. As shown in Fig.\ref{fig_slica}, the
black spots represent the silicon
atoms. The TEM images are processed to retain the distribution
information of the silicon atoms by thresholding the
grayness intensity distribution associated with each silicon atom and the associated spectral density $\hat{\psi}({\bf k})$ (where ${\bf
k}$ is the wave-vector) is then directly computed from the processed image following Ref.~\cite{ref41}
and shown in the inset of Fig.\ref{fig_slica}(b). The angularly averaged
$\hat{\psi}(k)$ (with $k = |{\bf k}|$) is shown Fig.\ref{fig_slica}(b).
The spectral density analysis used here is equivalent to the structure. It
can be seen that $\hat{\psi}({\bf k})$ is fully isotropic and the
scattering is completely suppressed at infinite wavelength, i.e.,
$\lim_{k\rightarrow 0}\hat{\psi}(k) = 0$ with $\hat{\psi}({\bf
k})\sim |{\bf k}|$ for small $k$ values, which indicates that the 2D
amorphous silica samples analyzed are class-II hyperuniform. The origin of hyperuniformity in these systems also result from the hyperuniformity-preserving nature of the Stone-Wales topological defects \cite{Zh20}.

The density of states (DOS) for both the hyperuniform amorphous silica systems and their crystalline counterpart were calculated at the DFT-PBE level of theory \cite{PhysRevLett.77.3865, chen2010systematically, li2016large}, see Fig.~\ref{fig_slica}(c) and (d), respectively. These calculations revealed the insulator nature of the 2D crystalline SiO$_2$, with a predicted band gap of 5.31 eV. By contrast, in the hyperuniform amorphous silica a small but finite number of states occupy the Fermi level of
the DHU structure, showing {\it metallic} behavior of the
electrons with a typical band gap of $\sim 0.05$ eV. This is
comparable to the thermal fluctuations at room temperature $\sim
0.025$ eV. The density $\rho$ of the electrons that contribute to the electrical conductivity of
the amorphous silica at room temperature was estimated as $\rho$ as 2.33 $\times 10^{12}$ cm$^{-2}$, by integrating the number
of states in the energies ranging from 25 meV (corresponding to
the thermal energy) to the Fermi level. This value belongs to
the category of ``high doping" regime (e.g., between 6.0 $\times 10^{11}$ and 9.2
$\times 10^{12}$ cm$^{-2}$) in common semi-metals \cite{yang2014chloride}.
In other words, the disordered hyperuniformity
fundamentally changes the electrical transport behavior of 2D
SiO$_2$, from an effective insulator at room temperature (as in
the crystalline form) to an effective metal (as in the disordered hyperuniform form).

The metallic behavior of disordered hyperuniform SiO$_2$ can also be seen from the distribution of charge densities within an energy window of 0.5 eV below the Fermi level (see Fig. \ref{fig_slica}(f)). For purpose of comparison, the charge density distribution for the crystalline structure within an energy window of 0.5 eV below the highest occupied molecular orbital (HOMO) level is shown in Fig. \ref{fig_slica}(e). The electrons in the crystalline structure fully occupy the valence
bands, but they cannot be thermally excited at room temperature to the conduction
bands due to the large band gap, leading to zero electrical
conductivity for pure 2D crystalline SiO$_2$. On the other hand, the number of valence
electrons in the disordered hyperuniform system in the same energy window is much less, resulting in a
low carrier density, but nevertheless, a non-zero conductivity. A
closer look at the slightly wider energy window associated with
lowest density of states (e.g. -2$\sim$0 eV) reveals that the
distribution of states still forms an almost continuous spectrum
of peaks, see the inset in Fig.~\ref{fig_slica}(d).
Figure~\ref{fig_slica}(f) reveals that the valence
electrons contributing to the conductivity originate from a small
portion of the Si and O atoms in the DHU system.


\subsection{Pentagonal 2D Materials}

\begin{figure}[htp]
\includegraphics[width=0.9\textwidth,keepaspectratio]{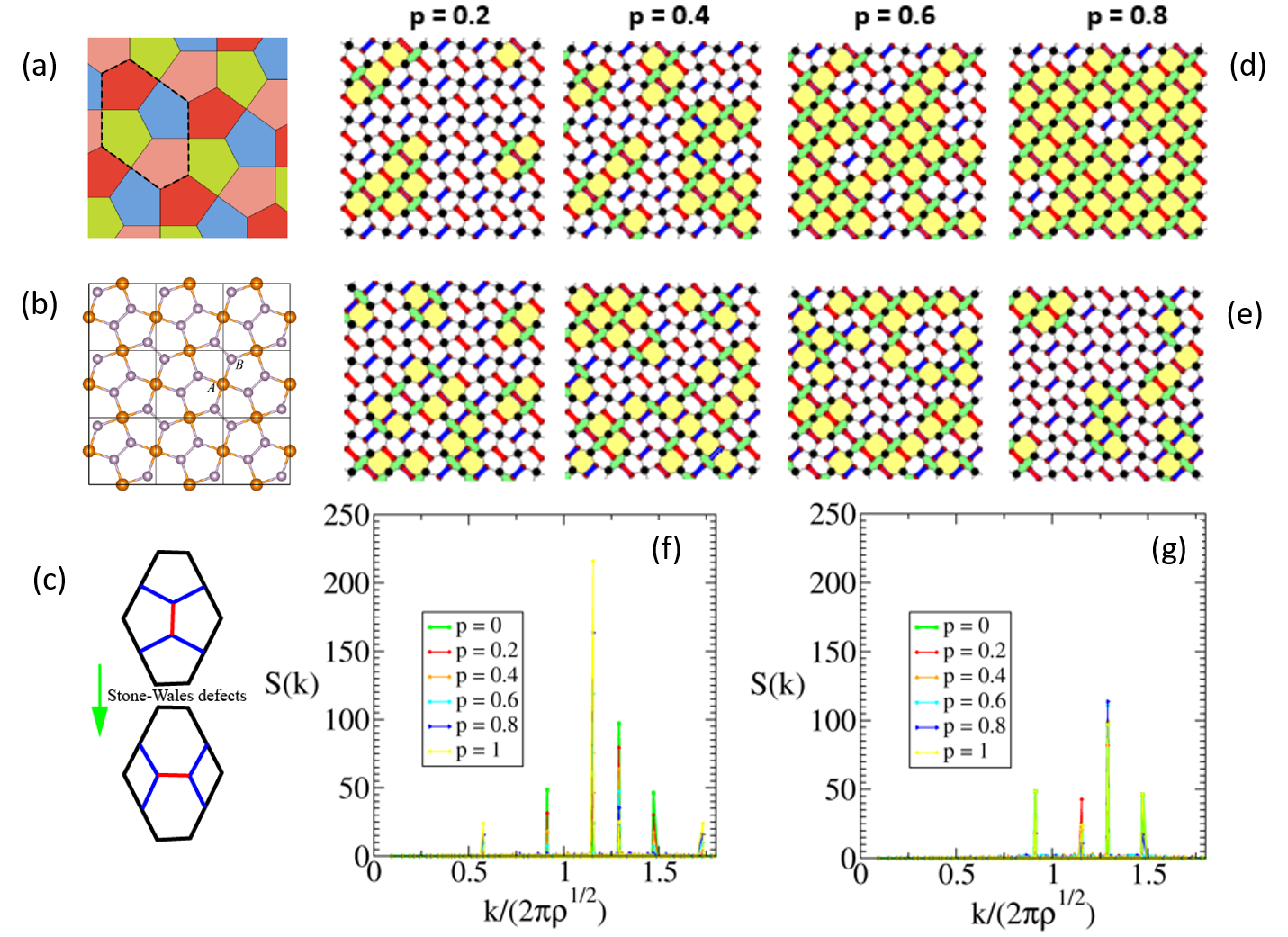}
\caption{Illustration of the Cairo tessellation formed with type II pentagons filling the plane (a), the pentagonal 2D $AB_2$ material mapped from the Cairo tessellation (b), and the Stone-Wales (SW) topological defect in the Cairo tessellation (c). (d) and (e) respectively illustrate two distinct topological pathways through which the pentagonal Cairo tiling (P5) respectively transforms into a crystalline rhombus-hexagon (C46) tiling (d) and random rhombus-pentagon-hexagon (R456) tilings (e) by continuously introducing the SW defects. The intermediate configurations along the P5-C46 pathway contain rhombi and hexagons whose orientations of the are all aligned in the same direction. (f) and (g) respectively show the structure factor $S(k)$ of the structures along the P5-C46 pathway (f) and P5-R456 pathway (g) as a function of the dimensionless wavenumber $k/(2\pi\rho^{1/2})$, where $\rho$ is the number density of the system.}
\label{fig_p501}
\end{figure}

Very recently, a class of pentagonal 2D materials have been discovered, which can be derived from the Cairo tessellation with type-II pentagons (see Fig. \ref{fig_p501}(a)), and realized by single-layer $AB_2$ pyrite structures (see {Fig. \ref{fig_p501}(b)}). An example of such 2D material is single-layer PtP$_2$, in which each fundamental cell contains two Pt atoms and four P atoms forming two pairs of pentagons touching through a vertex. Unlike their 2D hexagonal counterparts, 2D pentagonal materials, in particular, PdS$_2$ that have been successfully synthesized exhibit intrinsic in-plane anisotropy useful for various (e.g., thermoelectric) applications \cite{lu2020layer}.

\begin{figure}[ht]
\includegraphics[width=0.685\textwidth,keepaspectratio]{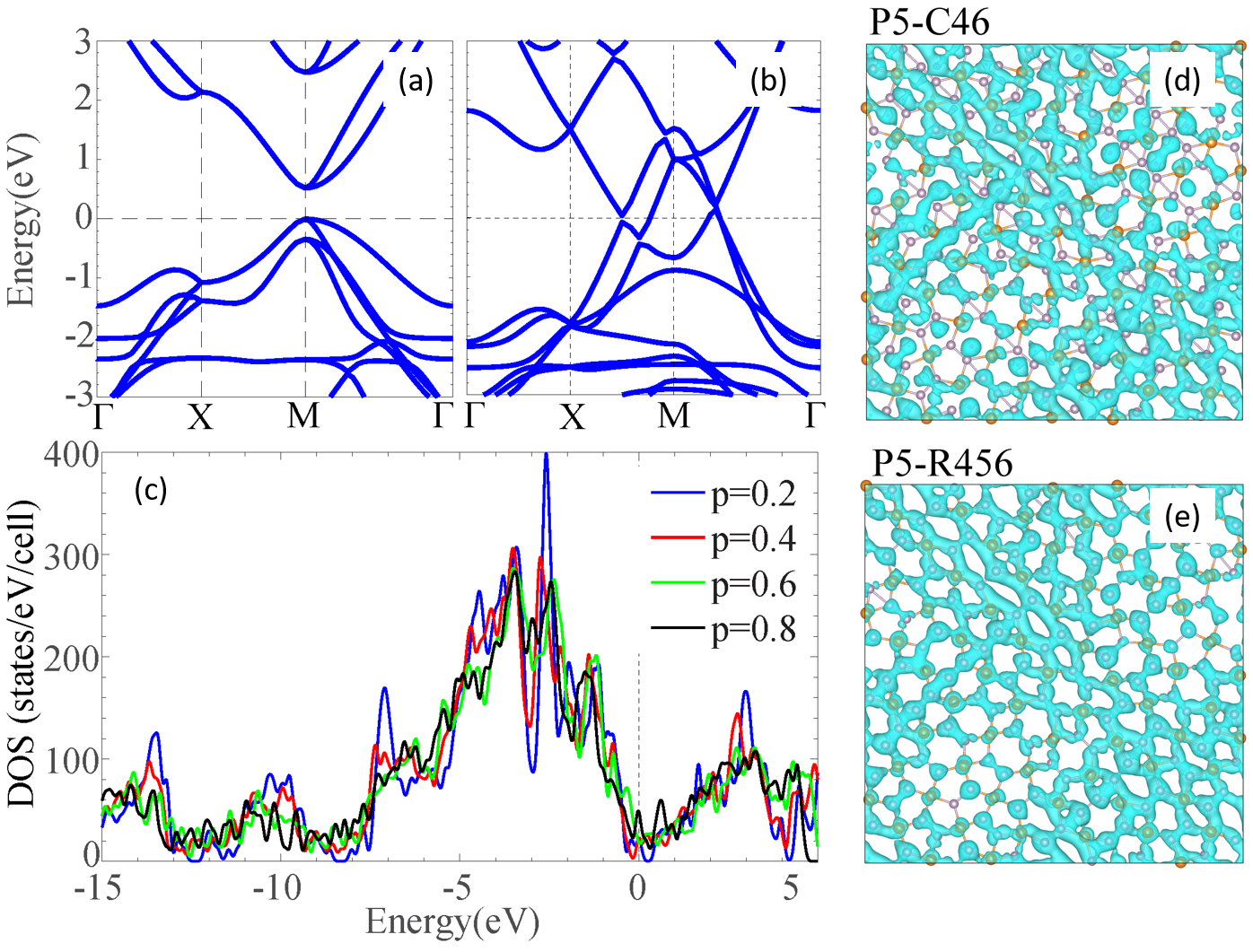}
\caption{Band structures of 2D PtP$_2$ with the (a) P5 \cite{PhysRevMaterials.2.114003}  and (b) C46 structures computed at the HSE06 level of theory. (c) Density of states of four intermediate structures ($p$ = 0.2, 0.4, 0.6, and 0.8) in the P5-C46 transition pathway calculated at the same level of theory. (d) and (e) respectively show partial charge densities that are $\pm 0.1$ eV about the Fermi level of two intermediate structures ($p$ = 0.5) in the P5-C46 (d) and P5-R456 (e) transition pathways calculated at the PBE level of theory. The isosurface value of the charge density plots is set to 0.0002$e$/$a_0^3$ (where $a_0$ is the Bohr radius).}
\label{fig_p502}
\end{figure}

Similar to the hexagonal 2D materials discussed above, Stone-Wales defects can be introduced into the crystalline pentagonal materials as a source of disorder \cite{Zh21}. A single SW defect in the fundamental cell of the Cairo tilling induces a local topological transformation which converts two pairs of pentagons into a pair of rhombi and a pair of hexagons touching through a common edge (See {Fig. \ref{fig_p501}(c)}). Unlike the SW defects in hexagonal 2D materials such as graphene, which cause distortions, the defects in pentagonal 2D materials preserve the shape and symmetry of the fundamental cell of P5 tiling and are associated with a minimal energy cost, making the intermediate R456 tilings realizable metastable states at room temperature \cite{Zh21}.


In Ref. \cite{Zh21}, two distinct topological pathways were reported through which the pentagonal Cairo tiling (P5) (i.e., single-layer $AB_2$ pyrite materials) respectively transforms into a crystalline rhombus-hexagon (C46) tiling and random rhombus-pentagon-hexagon (R456) tilings, by continuously introducing the Stone-Wales (SW) topological defects, see Fig. \ref{fig_p501}(d) and (e). It was found that these topological transformations are controlled by the orientation correlations among neighboring $B$-$B$ bonds, and exhibit a phenomenological analogy of the (anti)ferromagnetic to paramagnetic transition in two-state Ising systems.  Moreover, the intermediate structures along the two pathways are neither crystals nor quasicrystals, and yet these random tilings preserve hyperuniformity of the P5 or C46 crystal (i.e., the infinite-wavelength normalized density fluctuations are completely suppressed, see Fig. \ref{fig_p501}(f) and (g)), and can be viewed as 2D analogs of disordered Barlow packings in three dimensions.

The resulting 2D materials possess metal-like electronic properties, making them promising candidates for forming Schottky barriers with the semiconducting P5 material. Figure \ref{fig_p502}(a) and (b) compare the band structures of the 2D materials derived from the P5 and C46 structures. As reported in Ref. \cite{PhysRevMaterials.2.114003}, the P5 structure is semiconducting with a direct band gap of 0.52 eV. By contrast, the C46 structure is metallic with {\it Dirac-cone-like} dispersion in the conduction bands near the X point. This dispersion is caused by the presence of the hexagonal six-membered rings in the C46 structure. The density of states for four disordered intermediate structures with p = 0.2, 0.4, 0.6, and 0.8 along the P5-C46 transition pathway, calculated at the HSE06 level of theory, are shown in Fig. {\ref{fig_p502}(c)}. The closure of band gaps (marking the apparent continuous transition from semiconducting to metallic behaviors) as seen from the calculated DOS is already very evident in the structures with $p = 0.2$, and all band gaps are completely closed for structures associated with larger $p$. Additional calculations showed that the electron states around the Fermi level in these hyperuniform intermediate structures are itinerant rather than localized (see Fig. \ref{fig_p502}(d) and (e)), further indicating the metallic nature of these systems.

\subsection{Defected Two-Dimensional Transition Metal Dichalcogenides}

\begin{figure}[ht]
\includegraphics[width=0.925\textwidth,keepaspectratio]{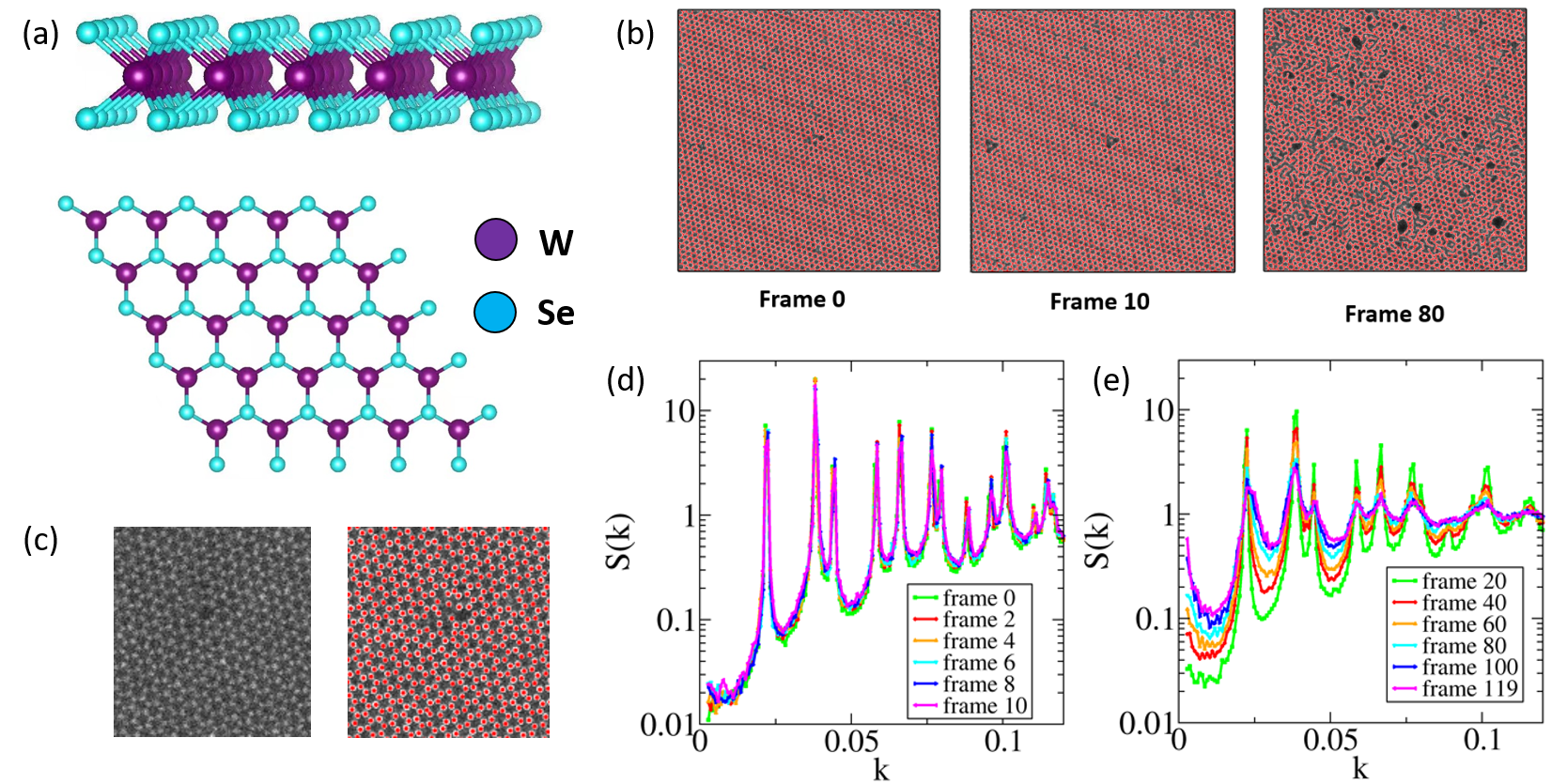}
\caption{(a) A schematic of the 2H-WSe$_2$ monolayer (upper panel) and its projection (lower panel) when seen from above. Note that the Te-doped 2H-WSe$_2$ monolayer (at low Te concentration can be viewed as effectively the 2H-WSe$_2$ monolayer), when projected from above, is mapped into a perfect honeycomb lattice, with each ``particle'' in the projected plane possessing three bonds. (b) A schematic illustrating the mapping from a raw image (top) of a defected monolayer crystalline transition metal dichalcogenide obtained using the annular dark-field scanning transmission electron microscopy (ADF-STEM) technique to the extracted atomic positions (bottom). (c) Chemical-bonding informed coordination networks overlaid with raw images of different frames during the evolution of the material system obtained using the ADF-STEM technique. Static structure factor $S(k)$ of the evolving system in nearly hyperuniform (d) and non-hyperuniform (e) regimes.}
\label{fig_tmdc}
\end{figure}

While SW defects are prevalent in graphene-like 2D materials, other types of defects such as chalcogen vacancies, metal vacancies and trefoil defects are dominant in monolayer transition metal dichalcogenides (TMDCs) such as MoS$_2$ and WSe$_{2}$ \cite{Li15, Le20} (see Fig. \ref{fig_tmdc}(a)). For example, Lin and coworkers \cite{Li15} have elaborated how the local structures evolve as various types of defects are introduced into MoS$_2$. Experimental studies \cite{Le19c, Le20b, Ko13} examining the evolution of structures as various types of defects are introduced into samples of TMDCs have also been reported.

In Ref. \cite{PhysRevB.103.224102}, a comprehensive characterization of the evolution of global structures as double chalcogen vacancies are gradually introduced into experimental samples of TMDCs (see Fig. \ref{fig_tmdc}(b)) was carried out within the hyperuniformity framework. Deep-learning algorithms were first employed to extract the atomic positions in a sequence of image frames obtained from ADF-STEM, as the scanning electron probe continuously introduces defects into a monolayer crystalline 2D transition metal dichalcogenide alloy, Te-doped 2H-WSe$_2$. The chemical-bonding informed coordination networks were subsequently constructed and refined for this evolving system (see Fig. \ref{fig_tmdc}(c)). In the perfect crystalline state, the coordination network is represented as a honeycomb network similar to crystalline graphene and silica.

Calculations of static structure factor $S(k)$ of the derived coordination networks indicated the systems are nearly hyperuniform at low defect concentrations (see Fig. \ref{fig_tmdc}(d)), and the hyperuniformity of the crystalline network is completely destroyed even when there is still a significant portion of crystalline sites in the system (specifically, less than $20 \%$ defects, see Fig. \ref{fig_tmdc}(e)). The effects of temperature on detecting hyperuniformity from experimental images were also investigated, which was found to be equivalent to inducing correlated displacement of lattice sites in a perfect crystalline honeycomb network. Moreover, it was shown that the structure factor of the system at relatively low concentration $p$ of double chalcogen vacancies assumes the analytical expression
\begin{equation}
S(k) \approx \frac{p}{2} + (1-\frac{p}{2})S_0(k),
\end{equation}
where $S_0(k)$ is the structure factor of the vacancy-free honeycomb network. These results reveal the level of degradation of hyperuniformity that one should expect due to the finite experimental measurement precision in STEM experiments. In addition, the analysis procedures can be readily adapted to characterize the structures of other ordered and disordered two-dimensional materials.






\section{Disordered Hyperuniform Quasi-1D Materials: Amorphous Nanotubes}

\begin{figure}[ht!]
\begin{center}
$\begin{array}{c}\\
\includegraphics[width=0.95\textwidth]{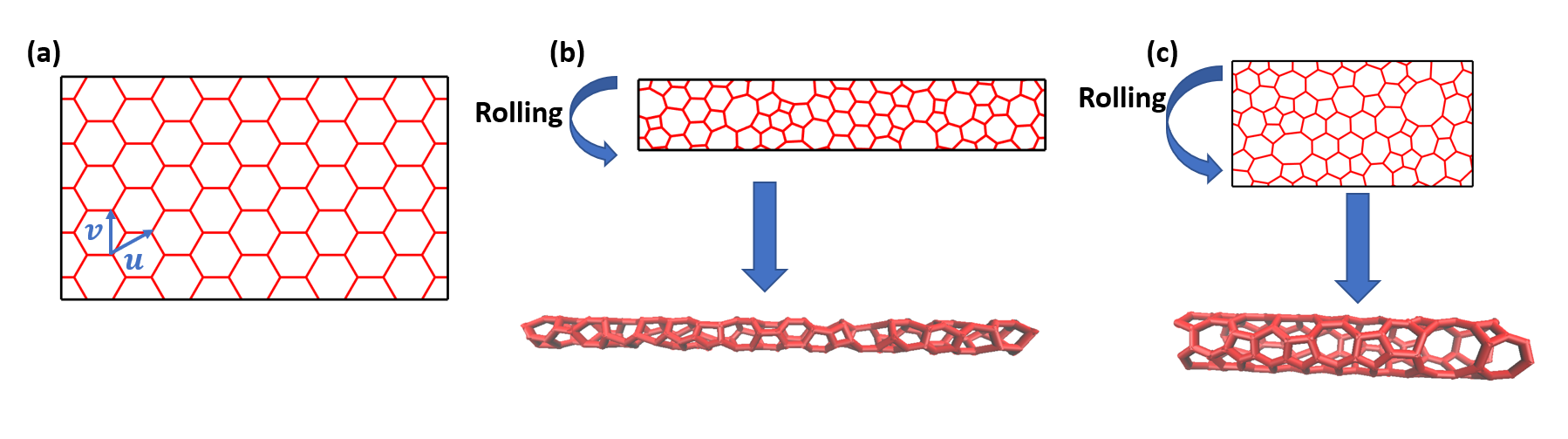}
\end{array}$
\end{center}
\caption{(a) Illustration of two linearly independent vectors ${\bf u}$ and ${\bf v}$ that form the basis to specify the rolling direction. A rolling vector $n \mathbf{u} + n \mathbf{v}$ uniquely determines the resulting nanotube structure from a given graphene sheet. (b) Illustration of a defected (3,0) zigzag nanotube formed by rolling up a graphene sheet with randomly distributed Stone-Wales defects along the rolling vector $3 \mathbf{u}$.  (c) Illustration of a defected (3,3) armchair nanotube formed by rolling up a graphene sheet with randomly distributed Stone-Wales defects along the rolling vector $3 \mathbf{u} + 3 \mathbf{v}$.} \label{fig_nano_1}
\end{figure}

Carbon nanotubes, a class of quasi-1D materials, can be conceptually constructed by rolling up one or multiple graphene sheets, and in this sense can be viewed as derivative structures of 2D materials. In particular, single-walled carbon nanotubes (SWNTs) can be formed by rolling up a long strip of a single graphene sheet along different directions \cite{Si01}. Conventionally, the type of SWNTs can be specified by a rolling vector $(n,m)$ (with $n>0$, $m\geq0$, and $n\geq m$) in the basis of two linearly independent vectors (e.g. ${\bf u}$ and ${\bf v}$ in Fig. \ref{fig_nano_1}(a)) that connect a carbon atom in the graphene sheet to either two of its nearest atoms with the same bond directions. Two most common types of SWNTs are: (i) zigzag nanotubes with $n>0$ and $m=0$ (see an example in Fig. \ref{fig_nano_1}(b)), and (ii) armchair nanotubes with $n=m$ (see an example in Fig. \ref{fig_nano_1}(c)). Multi-walled nanotubes (MWNTs) consisting of purely zigzag nanotubes, or purely armchair nanotubes, or a mixture of both are also of great interest.

\begin{figure}[ht!]
\begin{center}
$\begin{array}{c}\\
\includegraphics[width=0.98\textwidth]{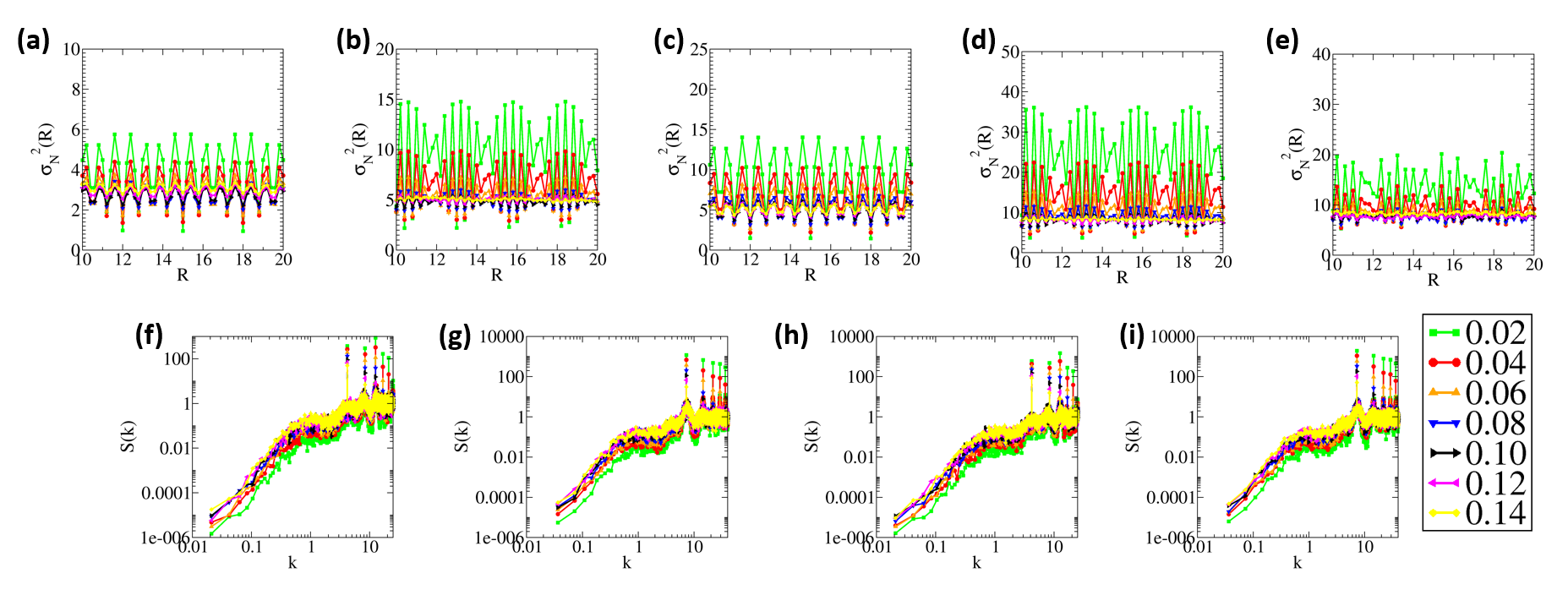}
\end{array}$
\end{center}
\caption{Local number variances $\sigma_N^2(R)$ and structure factor $S(k)$ of defected singled-walled and multi-walled nanotubes at different defect concentrations $p$. (a) $\sigma_N^2(R)$ of (5,0) zigzag nanotubes. (b) $\sigma_N^2(R)$ of (5,5) armchair nanotubes. (c) $\sigma_N^2(R)$ of double-walled nanotubes consisting of a (3,0) zigzag nanotube and a (5,0) zigzag nanotube. (d) $\sigma_N^2(R)$ of double-walled nanotubes consisting of a (3,3) armchair nanotube and a (5,5) armchair nanotube. (e) $\sigma_N^2(R)$ of double-walled nanotubes consisting of a (5,0) zigzag nanotube and a (5,5) armchair nanotube. (f) $S(k)$ of (5,0) zigzag nanotubes. (g) $S(k)$ of (5,5) armchair nanotubes. (h) $S(k)$ of double-walled nanotubes consisting of a (3,0) zigzag nanotube and a (5,0) zigzag nanotube. (i) $S(k)$ of double-walled nanotubes consisting of a (3,3) armchair nanotube and a (5,5) armchair nanotube. The results are all averaged over 10 configurations.} \label{fig_nano_2}
\end{figure}

\begin{figure}[ht!]
\begin{center}
$\begin{array}{c}\\
\includegraphics[width=0.95\textwidth]{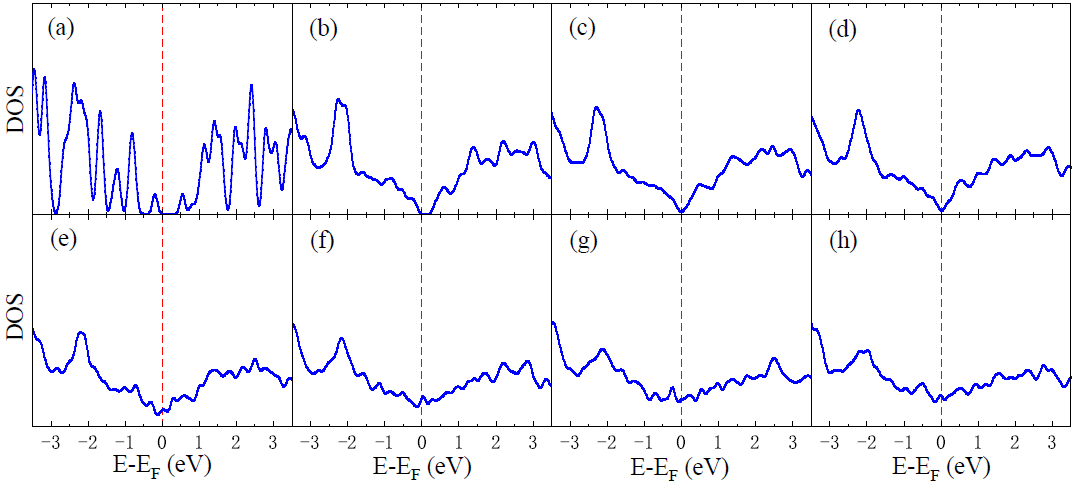}
\end{array}$
\end{center}
\caption{Density of states (DOS) of (10, 0) zigzag DHU carbon nanotubes computed by PBE functional at defect concentration $p$: (a) 0.00, (b) 0.01, (c) 0.0167, (d) 0.02, (e) 0.03, (f) 0.04, (g) 0.05, (h) 0.06.} \label{fig_nano_3}
\end{figure}

Similar to the cases of 2D materials, SW defects are commonly seen in carbon nanotubes (see Fig. \ref{fig_nano_1}(b) and (c) for illustrations of carbon nanotubes with randomly introduced SW defects), which generally can be introduced via high-energy radiations, during the synthesis of materials, or by applying strains \cite{Zh03}. Very recently, Chen and coauthors \cite{chen2022disordered} have systematically quantified the density fluctuations along the propagation direction in amorphous carbon nanotubes containing randomly distributed Stone-Wales defects. By effectively looking at 1D projections of the graphene sheets along the axial/propagation direction, they demonstrated that all of these amorphous nanotubes are hyperuniform, i.e., the infinite-wavelength density fluctuations of these systems are completely suppressed, regardless of the diameter, rolling axis, number of rolling sheets, and defect fraction of the nanotubes.

Specifically, the authors \cite{chen2022disordered} considered a generalized concept of hyperuniformity by taking into account that multiple points in the graphene sheets can be mapped to the same point in the projections. They generalized the definition of the structure factor $S({\bf k})$ to be:
\begin{equation}
    S({\bf k}) = \frac{1}{G} \left |{\sum_{j=1}^M g_j\exp(i {\bf k} \cdot
{\bf x}_j)}\right |^2 \quad ({\bf k} \neq {\bf 0}),
\end{equation}
where $M$ is the number of distinguishable points in the projections, the {\it multiplicity} $g_j$ is defined as the number of points in the higher-dimensional structures that are mapped to the given point ${\bf x}_j$ in the projections, and $G=\sum_{j=1}^M g_j$. Accordingly, the concept of $N(R)\equiv \langle N(R;{\bf x}_0)\rangle$ associated with $\sigma_N^2(R)$ was generalized to be:
\begin{equation}
    N(R;{\bf x}_0) = \sum_{j=1}^M g_j m({\bf x}_j-{\bf x}_0; R),
\end{equation}
where $\langle \cdots \rangle$ denotes ensemble average, $m({\bf x}-{\bf x}_0; R)$ is the indicator function of the observation window centered at ${\bf x}_0$ with radius $R$ and is defined as
\begin{equation}
m({\bf x; R}) = \left \{
\begin{array}{c@{\hspace{0.3cm}}c@{\hspace{0.3cm}}c}
1, & |{\bf x}| \leq R, &
\\ 0, & \textnormal{otherwise}. & \end{array} \right .
\end{equation}

It has been demonstrated \cite{chen2022disordered} that the local number variance $\sigma_{N}^2(R)$ of defect-free $(n,0)$ zigzag SWNTs or MWNTs consisting of $K$ zigzag nanotubes with $(n_1,0)$, $(n_2,0)$, $\cdots$, $(n_K,0)$ rolling vectors, respectively, can be determined analytically as
\begin{equation}
\sigma_{N}^2(R) = (n_1+n_2+\cdots+n_K)^2 \sigma_{N,D}^2(R),
\end{equation}
where $\sigma_{N,D}^2(R)$ is the local number variance of the projected periodic two-scale 1D point patterns \cite{To03} with $\zeta = \frac{1}{3}$, and $K=1$ is the special case for SWNTs. Here $\zeta$ is defined as the ratio of the nearest-neighbour distance over the length of the smallest repeating unit in the projected 1D point pattern. Similarly, the local number variance $\sigma_{N}^2(R)$ of defect-free $(n,n)$ armchair SWNTs or MWNTs consisting of $K$ armchair nanotubes with $(n_1,n_1)$, $(n_2,n_2)$, $\cdots$, $(n_K,n_K)$ rolling vectors, respectively, can be determined as
\begin{equation}
\sigma_{N}^2(R) = 4(n_1+n_2+\cdots+n_K)^2 \sigma_{N,S}^2(R),
\end{equation}
where $\sigma_{N,S}^2(R)$ is the local number variance of the projected periodic single-scale 1D pattern \cite{To03}, and $K=1$ is the special case for SWNTs. On the other hand, as SW defects are introduced into the nanotubes, the structures gradually transition into amorphous ones, which are reflected in their local number variance $\sigma_{N}^2(R)$ shown in Fig. \ref{fig_nano_2}. At low $p$, $\sigma_{N}^2(R)$ exhibits ``periodicity'' in the window radius $R$, indicating that the crystalline order is reminiscent in the systems; at large $p$, the oscillations of $\sigma_{N}^2(R)$ become much more damped as $R$ increases, suggesting the emergence of truly amorphous states. The variance $\sigma_{N}^2(R)$ of these SWNTs and MWNTs fluctuate around certain constants as $R$ increases at all investigated $p$, indicating that these structures are class-I hyperuniform. The results of ensemble-averaged $S(k)$ of the aforementioned zigzag and armchair nanotubes are also shown in Fig. \ref{fig_nano_2}, which all decreases to zero as $k$ goes to zero, regardless of defect concentration $p$. These results further confirm the hyperuniformity of these nanotubes, and are consistent with the results of $\sigma_{N}^2(R)$.

Interestingly, MWNTs consisting of both zigzag and armchair nanotubes exhibit different behaviors in their density fluctuations from SWNTs or MWNTs consisting of purely zigzag or armchair nanotubes \cite{chen2022disordered}. For example, as shown in Fig. \ref{fig_nano_2}(e), $\sigma_{N}^2(R)$ of MWNT consisting of a (5,0)-zigzag nanotube and a (5,5)-armchair nanotube exhibit no ``periodicity'' in the window radius $R$ even at low $p$. This is a direct result of the fact that the length of the smallest repeating unit in a defect-free (5,0)-zigzag nanotube and that of the smallest repeating unit in a defect-free (5,5)-armchair nanotube do not have an integer common multiple, and the 1D projections of the two nanotubes collectively onto the cylinder axis are no longer periodic.

The electronic structures of DHU nanotubes with SW defects have also been investigated. For example, as shown in Fig.~\ref{fig_nano_3} \cite{chen2022disordered}, the density of states (DOSs) of (10, 0) zigzag DHU carbon nanotubes at different $p$ as well as the DOS of defect-free carbon nanotubes. While it is well known \cite{Az10, Pa13} that (10,0) nanotubes possess a well-defined band gap at Fermi level, increasing disorder in the DHU system (i.e., increasing the amount of SW defects) results in two observed effects on the computed DOS \cite{chen2022disordered, Cr97}: (i) closure of the band gap at Fermi level and (ii) broadening and flattening of the DOS. This trend is also similar to the observations in DHU 2D materials \cite{Zh20, Ch21}. Specifically, the band gap is closed at $p = 0.0167$, suggesting the presence of the semiconductor-to-metal transition around this defect concentration. Moreover, as $p$ increases the DOS becomes more and more extended, converging to a metallic characteristic.














\section{Outlook on DHU Quantum Materials}


Although still in its infancy, the discoveries of disordered hyperuniform (DHU) solid-state materials, in particular the 2D and quasi-1D materials, shed light on the feasibility of identifying and engineering novel DHU quantum materials and DHU quantum states. There are a variety of potential device applications for disordered hyperuniform solid state materials. For example, quantum emitters that are capable of generating single photons have been proposed, which could be realized by introducing defects into two-dimensional (2D) materials such as hexagonal boron nitride ($h$-BN) \cite{tran2016quantum}. These materials are promising platforms to realize quantum bits (``qubits'') at room temperature \cite{Li19}. It is noteworthy that quantum decoherence \cite{Li19}, a process in which the state of the qubit loses it intrinsic quantum properties via interactions with the environment. This inherently changes the state leading to the quantum information loss and ultimately translates to loss of quantum operation fidelity and ultimately error generation. Naturally, the routes to control and suppress quantum decoherence constitute the crucial issue that needs to be addressed in the material research for quantum information science and they are critically determine the applicability of the current and future quantum computers \cite{Li19}.

In this context DHU materials offer a potentially unique platform for realizing decoherence protection by restricting the available collective lattice vibrational modes (i.e., presence of phononic gaps) and electron localization that limits long-range couplings. To this end, a comprehensive understanding of DHU in the context of quantum correlations and excited states is required, which is currently lacking. Many fundamental questions need to be addressed: Can addressable local excitations (such as quantum defects) be hosted by DHU systems suppressing coupling to the lattice excitations (phonons)?  Similarly, can such decoupling be realized in a fully quantum regime similar to driven decoupling that increases the coherence times in solid state qubits, such that spatial (and spatiotemporal) noise is employed instead of the time domain environmental noise for decoupling \cite{joos2022protecting}? How to control and engineer such states? How to properly generalize the notion of DHU to quantum states? How do DHU states emerge in single or correlated electron systems? How can the properties of individual DHU materials be leveraged to create interfaced DHU quantum materials?




Addressing these questions requires tackling the fundamental underlying physics of DHU systems and (quantum) states. This knowledge would feed into a design of DHU quantum materials targeting suppression or enhancement of particular types of couplings, mitigating some of the outstanding problems associated with quantum information manipulation and storage. In practice, the various types of correlated disorder intrinsic to 2D quantum material systems need to be revealed, quantified and categorized within the hyperuniformity framework that is distinctly different from the notion of disorder treated by currently existing theories and models. The distinguishable nature of the variety of hyperuniform disorder, in particular the unique quasi-long range (QLR) correlations, and their effects on quantum decoherence should be investigated and elucidated. Here, first principles computations and materials simulations are indispensable for novel compound discovery and design.



In particular, the nuclear, electronic, and their coupled degrees of freedom are clearly a fruitful area for exploration. We envision the following three specific research directions that will advance our fundamental understanding of the novel quantum DHU states and the fundamental roles played by the defects and disorder in quantum materials:


{\bf (i) Understanding the electronic structure in DHU systems and exploration of strong electron-electron correlations that result in new DHU quantum states.} Generalization of the notion of DHU states to quantum systems could be achieved by exploring two superficially similar yet very distinct realizations of quantum hyperuniformity in electronic states: 1) those supported by a DHU atomic lattice and (2) those emerging from the mutual quantum interactions regardless of the atomic geometric arrangement. Realizing quasiparticle states for which the long-range oscillations are effectively near-sighted are of immense potential application for protecting quantum states from dielectric noise, which is typically long-ranged. We surmise that this investigation will provide a unified picture of how electronic states may, under favorbale conditions, be protected against decoherence through favorable (de)coupling of the electronic fluctuations.

{\bf (ii) DHU-driven phonon decoherence protection of quantum defect states.} DHU materials behave as a crystalline solid in the quasi-long range and an amorphous material in the short range. If such systems harbor strongly localized electronic states, they will naturally couple to the lattice only locally (in analogy to the nearsightedness mentioned in the preceding section) and hence possibly exhibit advantages over other types of solid-state qubits. In the idealized extreme case, such quantum states will behave like behaving like trapped atoms/ions, are completely isolated from the external environment and thereby maximizing the coherence time. Further, in contrast to delocalized electronic states in a periodic system, which enhance the entanglements among qubits. DHU platform based qubits are positioned between the two extremes and likely to share both advantages, leading to ideal optimal coherent time, strong inter-qubit coupling, and maximally compact physical platform. This context creates a clear and strong incentive to design DHU systems with optimal phonon degrees of freedom and electron-phonon couplings to realize effective decoherence protection. Note that the phononically gaped systems would also constitute an ideal encapsulation layer for various quantum materials protecting them from environmental couplings.



While the DHU order is distinct from ideally periodic and completely random arrangements, it is not simply an interpolation between the two ``conventional'' states of matter. One of the most promising areas of DHU application is the realization of wide phononic gaps and phonon dispersion narrowed to localized modes. Both effects hold potential to decouple the electronic and vibrational degrees of freedom. Such a scenario is particularly appealing for cases when the DHU system ``embeds'' a localized and addressable quantum state (i.e., a qubit in practice). To an extent, this hypothesized mechanism can be paralleled with microscale acoustic optomechanical (de)coupling, which has been explored on multiple fronts in quantum transductions \cite{noguchi2017qubit, mirhosseini2020superconducting}. Here, however, the interactions with mechanical (acoustic) modes are governed by patterning the systems at the atomistic level, i.e., by introducing additional structural defects toward the DHU limit.


{\bf (iii) Emergent DHU ordering at the interfaces of 2D DHU heterostructures.} The DHU state of matter has been associated with a single 2D DHU quantum material so far. Yet, the fundamental concept is general and relates, in principle, to DHU heterostructures. Conventional 2D heterostructures have important quantum information science and engineering implications. For example, twisting one sheet of a 2D material (e.g., graphene) with respect to another causes an angle-dependent Moire superlattice generally with an extremely large unit cell and localized electronic states near the Fermi level exhibiting all hallmarks of strong correlations: when applying a strong electric field, bilayer graphene exhibits an insulating behavior at half-filling of the localized Moire states (paralleled with Mott insulator physics) and a related superconducting regime upon doping \cite{wang2021spin}. This behavior occurs at a small twist ``magic'' angle of 1.1 degree, which leads to strong coupling between two graphene monolayers and formation of flat (localized) states. The behavior of electrons at other superlattice arrangements can drastically differ, even in the large inter-layer coupling limit (e.g., at high pressures) \cite{romanova2022stochastic}. The mechanisms behind unconventional superconductivity is not fully understood. These and similar pioneering experiments on van der Waals bonded multi-layer systems with mutual rotational "mis-arrangement" have led to a burgeoning field called ``twistronics'', which has recently morphed into a new playground for exploring intriguing physics owing to strong electron-electron correlations. Given the intriguing quantum phenomena in conventional 2D heterostructures, it is keen to discover 2D DHU heterostructures that could serve as a novel platform to explore exotic quantum effects.

\begin{acknowledgments}
The authors are very grateful to Dr. Sef Tongay for his kind invitation and inspiration for this focused review, and to Dr. Yu Zheng, Dr. Lei Liu, Dr. Duo Wang, Chia-Hao Lee, Dr. Sangmin Kang, Dr. Wenjuan Zhu, Xinyu Jiang, Yu Liu, Justine Ilyssa Vidallon for their contributions to the work on disordered solid-state materials discussed in this review, and to Dr. Sal Torquato, Dr. Ge Zhang, and Dr. Jaeuk Kim for very helpful discussions. Y. J. was supported by the Army Research Office under Cooperative Agreement Number W911NF-22-2-0103. P. H. was supported by U.S. Department of Energy, Office of Science, Office of Basic Energy Sciences, Division of Materials Sciences and Engineering under award number DE-SC0020190
\end{acknowledgments}

\end{document}